\documentclass[twocolumn,letterpaper,aps,prd,superscriptaddress,showpacs,nofootinbib,floatfix]{revtex4}

\usepackage{graphicx}	

\usepackage{xspace}	
\usepackage{amsmath}
\usepackage{multirow}

\newcommand{\ppup}{\mbox{$p^{\uparrow}$$+$$p$}\xspace}

\newcommand{\minbias}{minimum bias\xspace}
\newcommand{\Minbias}{Minimum bias\xspace}
\newcommand{\highpt}{4$\times$4B\xspace}
\newcommand{\tile}{4$\times$4\xspace}
\newcommand{\pT}{$p_{T}$\xspace}

\begin{document}

\title{Cross section and transverse single-spin asymmetry of $\eta$ 
mesons in $p^{\uparrow}+p$ collisions at $\sqrt{s}=200$ GeV at forward rapidity}

\newcommand{\abilene}{Abilene Christian University, Abilene, Texas 79699, USA}
\newcommand{\augie}{Department of Physics, Augustana College, Sioux Falls, South Dakota 57197, USA}
\newcommand{\banaras}{Department of Physics, Banaras Hindu University, Varanasi 221005, India}
\newcommand{\barc}{Bhabha Atomic Research Centre, Bombay 400 085, India}
\newcommand{\baruch}{Baruch College, City University of New York, New York, New York, 10010 USA}
\newcommand{\bnlcoll}{Collider-Accelerator Department, Brookhaven National Laboratory, Upton, New York 11973-5000, USA}
\newcommand{\bnlphys}{Physics Department, Brookhaven National Laboratory, Upton, New York 11973-5000, USA}
\newcommand{\caucr}{University of California - Riverside, Riverside, California 92521, USA}
\newcommand{\charlesczech}{Charles University, Ovocn\'{y} trh 5, Praha 1, 116 36, Prague, Czech Republic}
\newcommand{\chonbuk}{Chonbuk National University, Jeonju, 561-756, Korea}
\newcommand{\ciae}{Science and Technology on Nuclear Data Laboratory, China Institute of Atomic Energy, Beijing 102413, People's~Republic~of~China}
\newcommand{\cns}{Center for Nuclear Study, Graduate School of Science, University of Tokyo, 7-3-1 Hongo, Bunkyo, Tokyo 113-0033, Japan}
\newcommand{\colorado}{University of Colorado, Boulder, Colorado 80309, USA}
\newcommand{\columbia}{Columbia University, New York, New York 10027 and Nevis Laboratories, Irvington, New York 10533, USA}
\newcommand{\czechtech}{Czech Technical University, Zikova 4, 166 36 Prague 6, Czech Republic}
\newcommand{\dapnia}{Dapnia, CEA Saclay, F-91191, Gif-sur-Yvette, France}
\newcommand{\elte}{ELTE, E{\"o}tv{\"o}s Lor{\'a}nd University, H - 1117 Budapest, P\'azmany P\'eter s\'et\'any 1/A, Hungary}
\newcommand{\ewha}{Ewha Womans University, Seoul 120-750, Korea}
\newcommand{\fit}{Florida Institute of Technology, Melbourne, Florida 32901, USA}
\newcommand{\fsu}{Florida State University, Tallahassee, Florida 32306, USA}
\newcommand{\gsu}{Georgia State University, Atlanta, Georgia 30303, USA}
\newcommand{\hanyang}{Hanyang University, Seoul 133-792, Korea}
\newcommand{\hiroshima}{Hiroshima University, Kagamiyama, Higashi-Hiroshima 739-8526, Japan}
\newcommand{\howard}{Department of Physics and Astronomy, Howard University, Washington, DC 20059, USA}
\newcommand{\ihepprot}{IHEP Protvino, State Research Center of Russian Federation, Institute for High Energy Physics, Protvino, 142281, Russia}
\newcommand{\illuiuc}{University of Illinois at Urbana-Champaign, Urbana, Illinois 61801, USA}
\newcommand{\inrras}{Institute for Nuclear Research of the Russian Academy of Sciences, prospekt 60-letiya Oktyabrya 7a, Moscow 117312, Russia}
\newcommand{\instpasczech}{Institute of Physics, Academy of Sciences of the Czech Republic, Na Slovance 2, 182 21 Prague 8, Czech Republic}
\newcommand{\isu}{Iowa State University, Ames, Iowa 50011, USA}
\newcommand{\jaea}{Advanced Science Research Center, Japan Atomic Energy Agency, 2-4 Shirakata Shirane, Tokai-mura, Naka-gun, Ibaraki-ken 319-1195, Japan}
\newcommand{\jyvaskyla}{Helsinki Institute of Physics and University of Jyv{\"a}skyl{\"a}, P.O.Box 35, FI-40014 Jyv{\"a}skyl{\"a}, Finland}
\newcommand{\kek}{KEK, High Energy Accelerator Research Organization, Tsukuba, Ibaraki 305-0801, Japan}
\newcommand{\korea}{Korea University, Seoul, 136-701, Korea}
\newcommand{\kurchatov}{Russian Research Center ``Kurchatov Institute," Moscow, 123098 Russia}
\newcommand{\kyoto}{Kyoto University, Kyoto 606-8502, Japan}
\newcommand{\labllr}{Laboratoire Leprince-Ringuet, Ecole Polytechnique, CNRS-IN2P3, Route de Saclay, F-91128, Palaiseau, France}
\newcommand{\lahorelums}{Physics Department, Lahore University of Management Sciences, Lahore 54792, Pakistan}
\newcommand{\lawllnl}{Lawrence Livermore National Laboratory, Livermore, California 94550, USA}
\newcommand{\losalamos}{Los Alamos National Laboratory, Los Alamos, New Mexico 87545, USA}
\newcommand{\lpc}{LPC, Universit{\'e} Blaise Pascal, CNRS-IN2P3, Clermont-Fd, 63177 Aubiere Cedex, France}
\newcommand{\lund}{Department of Physics, Lund University, Box 118, SE-221 00 Lund, Sweden}
\newcommand{\maryland}{University of Maryland, College Park, Maryland 20742, USA}
\newcommand{\mass}{Department of Physics, University of Massachusetts, Amherst, Massachusetts 01003-9337, USA }
\newcommand{\michigan}{Department of Physics, University of Michigan, Ann Arbor, Michigan 48109-1040, USA}
\newcommand{\muenster}{Institut fur Kernphysik, University of Muenster, D-48149 Muenster, Germany}
\newcommand{\muhlenberg}{Muhlenberg College, Allentown, Pennsylvania 18104-5586, USA}
\newcommand{\myongji}{Myongji University, Yongin, Kyonggido 449-728, Korea}
\newcommand{\nagasaki}{Nagasaki Institute of Applied Science, Nagasaki-shi, Nagasaki 851-0193, Japan}
\newcommand{\newmex}{University of New Mexico, Albuquerque, New Mexico 87131, USA }
\newcommand{\nmsu}{New Mexico State University, Las Cruces, New Mexico 88003, USA}
\newcommand{\ohio}{Department of Physics and Astronomy, Ohio University, Athens, Ohio 45701, USA}
\newcommand{\ornl}{Oak Ridge National Laboratory, Oak Ridge, Tennessee 37831, USA}
\newcommand{\orsay}{IPN-Orsay, Universite Paris Sud, CNRS-IN2P3, BP1, F-91406, Orsay, France}
\newcommand{\peking}{Peking University, Beijing 100871, People's~Republic~of~China}
\newcommand{\pnpi}{PNPI, Petersburg Nuclear Physics Institute, Gatchina, Leningrad Region, 188300, Russia}
\newcommand{\riken}{RIKEN Nishina Center for Accelerator-Based Science, Wako, Saitama 351-0198, Japan}
\newcommand{\rikjrbrc}{RIKEN BNL Research Center, Brookhaven National Laboratory, Upton, New York 11973-5000, USA}
\newcommand{\rikkyo}{Physics Department, Rikkyo University, 3-34-1 Nishi-Ikebukuro, Toshima, Tokyo 171-8501, Japan}
\newcommand{\saispbstu}{Saint Petersburg State Polytechnic University, St. Petersburg, 195251 Russia}
\newcommand{\saopaulo}{Universidade de S{\~a}o Paulo, Instituto de F\'{\i}sica, Caixa Postal 66318, S{\~a}o Paulo CEP05315-970, Brazil}
\newcommand{\seoulnat}{Department of Physics and Astronomy, Seoul National University, Seoul 151-742, Korea}
\newcommand{\stonybrkc}{Chemistry Department, Stony Brook University, SUNY, Stony Brook, New York 11794-3400, USA}
\newcommand{\stonycrkp}{Department of Physics and Astronomy, Stony Brook University, SUNY, Stony Brook, New York 11794-3800, USA}
\newcommand{\tenn}{University of Tennessee, Knoxville, Tennessee 37996, USA}
\newcommand{\titech}{Department of Physics, Tokyo Institute of Technology, Oh-okayama, Meguro, Tokyo 152-8551, Japan}
\newcommand{\tsukuba}{Institute of Physics, University of Tsukuba, Tsukuba, Ibaraki 305, Japan}
\newcommand{\vandy}{Vanderbilt University, Nashville, Tennessee 37235, USA}
\newcommand{\waseda}{Waseda University, Advanced Research Institute for Science and Engineering, 17 Kikui-cho, Shinjuku-ku, Tokyo 162-0044, Japan}
\newcommand{\weizmann}{Weizmann Institute, Rehovot 76100, Israel}
\newcommand{\wigner}{Institute for Particle and Nuclear Physics, Wigner Research Centre for Physics, Hungarian Academy of Sciences (Wigner RCP, RMKI) H-1525 Budapest 114, POBox 49, Budapest, Hungary}
\newcommand{\yonsei}{Yonsei University, IPAP, Seoul 120-749, Korea}
\newcommand{\zagreb}{University of Zagreb, Faculty of Science, Department of Physics, Bijeni\v{c}ka 32, HR-10002 Zagreb, Croatia}
\affiliation{\abilene}
\affiliation{\augie}
\affiliation{\banaras}
\affiliation{\barc}
\affiliation{\baruch}
\affiliation{\bnlcoll}
\affiliation{\bnlphys}
\affiliation{\caucr}
\affiliation{\charlesczech}
\affiliation{\chonbuk}
\affiliation{\ciae}
\affiliation{\cns}
\affiliation{\colorado}
\affiliation{\columbia}
\affiliation{\czechtech}
\affiliation{\dapnia}
\affiliation{\elte}
\affiliation{\ewha}
\affiliation{\fit}
\affiliation{\fsu}
\affiliation{\gsu}
\affiliation{\hanyang}
\affiliation{\hiroshima}
\affiliation{\howard}
\affiliation{\ihepprot}
\affiliation{\illuiuc}
\affiliation{\inrras}
\affiliation{\instpasczech}
\affiliation{\isu}
\affiliation{\jaea}
\affiliation{\jyvaskyla}
\affiliation{\kek}
\affiliation{\korea}
\affiliation{\kurchatov}
\affiliation{\kyoto}
\affiliation{\labllr}
\affiliation{\lahorelums}
\affiliation{\lawllnl}
\affiliation{\losalamos}
\affiliation{\lpc}
\affiliation{\lund}
\affiliation{\maryland}
\affiliation{\mass}
\affiliation{\michigan}
\affiliation{\muenster}
\affiliation{\muhlenberg}
\affiliation{\myongji}
\affiliation{\nagasaki}
\affiliation{\newmex}
\affiliation{\nmsu}
\affiliation{\ohio}
\affiliation{\ornl}
\affiliation{\orsay}
\affiliation{\peking}
\affiliation{\pnpi}
\affiliation{\riken}
\affiliation{\rikjrbrc}
\affiliation{\rikkyo}
\affiliation{\saispbstu}
\affiliation{\saopaulo}
\affiliation{\seoulnat}
\affiliation{\stonybrkc}
\affiliation{\stonycrkp}
\affiliation{\tenn}
\affiliation{\titech}
\affiliation{\tsukuba}
\affiliation{\vandy}
\affiliation{\waseda}
\affiliation{\weizmann}
\affiliation{\wigner}
\affiliation{\yonsei}
\affiliation{\zagreb}
\author{A.~Adare} \affiliation{\colorado}
\author{C.~Aidala} \affiliation{\mass} \affiliation{\michigan}
\author{N.N.~Ajitanand} \affiliation{\stonybrkc}
\author{Y.~Akiba} \affiliation{\riken} \affiliation{\rikjrbrc}
\author{R.~Akimoto} \affiliation{\cns}
\author{H.~Al-Bataineh} \affiliation{\nmsu}
\author{J.~Alexander} \affiliation{\stonybrkc}
\author{M.~Alfred} \affiliation{\howard}
\author{A.~Angerami} \affiliation{\columbia}
\author{K.~Aoki} \affiliation{\kyoto} \affiliation{\riken}
% ...
\collaboration{PHENIX Collaboration} \noaffiliation

\date{\today}


\begin{abstract}


We present a measurement of the cross section and transverse single-spin
asymmetry ($A_N$) for $\eta$ mesons at large pseudorapidity from
$\sqrt{s}=200$~GeV $p^{\uparrow}+p$ collisions. The measured cross section for
$0.5<p_T<5.0$~GeV/$c$ and $3.0<|\eta|<3.8$ is well described by a
next-to-leading-order perturbative-quantum-chromodynamics calculation. The
asymmetries $A_N$ have been measured as a function of Feynman-$x$ ($x_F$) from
$0.2<|x_{F}|<0.7$, as well as transverse momentum ($p_T$) from
$1.0<p_T<4.5$~GeV/$c$.  The asymmetry averaged over positive $x_F$ is
$\langle{A_{N}}\rangle=0.061{\pm}0.014$. The results are consistent with prior
transverse single-spin measurements of forward $\eta$ and $\pi^{0}$ mesons at
various energies in overlapping $x_F$ ranges. Comparison of different particle
species can help to determine the origin of the large observed asymmetries in
$p^{\uparrow}+p$ collisions.

\end{abstract}


\maketitle

\section{Introduction}

Since the proton's magnetic moment was revealed to be 2.79 times the size of 
the Dirac magnetic moment~\cite{magmoment}, studying the internal 
structure of the proton has been a vibrant field of physics research. 
Early deep-inelastic electron-nucleon scattering (DIS) experiments found 
that leptons were elastically scattered off of partons~\cite{dis0, dis1, 
dis2}, and further measurements have led to detailed understanding of the parton
distribution functions (PDFs) that can be used to describe the collinear
quark and gluon structure of the nucleon.  At leading order in a 
perturbative quantum chromodynamics (pQCD) expansion in the strong 
coupling $\alpha_s$, PDF $f(x)$ represents the probability of a parton of 
flavor $f$ carrying momentum fraction $x$ of the total proton momentum.  
The PDFs themselves are nonperturbative and cannot be calculated directly 
in pQCD; they must instead be extracted from experimental measurements.  
From the development of QCD until the 1990s, experimental and theoretical 
studies focused on the one-dimensional momentum structure of the nucleon, 
in which the partons are treated as moving collinearly with the parent 
nucleon.  Over the last two decades, a variety of theoretical and 
experimental tools have been developed to study other aspects of nucleon 
structure, including parton transverse dynamics within the nucleon.  The 
measurement of transverse single spin asymmetries (SSAs) provides one 
window into dynamical spin-momentum correlations both in QCD bound states 
and in the process of partonic hadronization.

Leading-twist pQCD calculations predict very small transverse single spin asymmetries,
less than $\mathcal{O}(10^{-4})$ at high-$p_{T}$
($p_{T}$\,$>$\,few GeV/$c$)~\cite{ansmall_pqcd}.  However, strikingly large
transverse SSAs, up to $\sim 40$\%, have been measured at forward rapidity for
hadrons produced from transversely polarized proton collisions ($p^{\uparrow}+p
\rightarrow h + X$), revealing significant spin-momentum correlations in the
nonperturbative structure of the proton.  These asymmetries have been observed
for collision energies ranging from
$\sqrt{s}$\,=\,4.9~to~500\,GeV~\cite{an_underpqcd0, an_underpqcd1,
an_underpqcd2, e704_0, e704_1, e704eta, anbrahms0, anstar0, stareta0,
Adare:2013ekj, Heppelmann:2013ewa} and for hadron transverse momenta ($p_T$) up
to 7\,GeV/$c$~\cite{Heppelmann:2013ewa}.  The persistence of transverse SSAs
into kinematic regimes where pQCD is applicable offers an opportunity to
describe this nonperturbative behavior in terms of well-defined functions using
the framework of pQCD. At midrapidity, no significant $A_N$ has been observed
\cite{Adler:2005in, Adare:2013ekj}.

Multiple approaches have been proposed to describe the large transverse SSAs
observed in hadronic reactions.  Transverse-momentum-dependent (TMD) PDFs
include explicit dependence not only on the partonic collinear momentum
fraction but also on the partonic transverse momentum ($k_T$) within the
nucleon.  Similarly, TMD fragmentation functions (FFs) depend on both the
collinear momentum fraction of the scattered parton acquired by the produced
hadron as well as the transverse momentum of the hadron with respect to the
direction of the scattered parton.  Reactions involving scattering of a proton
with its spin perpendicular to its momentum inducing the production of a hadron
can provide sensitivity to both initial-state (PDF) and final-state (FF)
effects.  

Sivers proposed a TMD PDF~\cite{sivers0, sivers1} as a possible origin of the
large observed transverse SSAs, corresponding to a correlation between the spin
of the proton and the transverse momentum of the quarks.  Semi-inclusive DIS
experiments have found evidence for a nonzero Sivers TMD
PDF~\cite{Airapetian:2004tw, Airapetian:2009ae, Alekseev:2010rw,
Adolph:2012sp}.  Collins alternatively proposed a TMD FF~\cite{Collins:1992kk}
that generates transverse SSAs, corresponding to a correlation between the
(transverse) polarization of a scattered quark and the angular distribution of
pions in the quark jet.  The outgoing quarks in \ppup collisions will have a
net transverse polarization if the transversity distribution in the proton is
nonzero.  Electron-positron annihilation, as well as semi-inclusive DIS
measurements, have now found evidence for a nonzero Collins TMD FF as well as a
nonzero transversity distribution~\cite{Airapetian:2004tw, Abe:2005zx,
Seidl:2008xc, Lees:2013yha, Alekseev:2010rw, Airapetian:2010ds, Adolph:2012sn}.
All these results indicate that there are sizable spin-momentum correlation
effects in QCD bound states as well as in the process of hadronization.

While these spin-momentum correlations are present in the proton and in the
process of hadronization, inclusive hadron production in \ppup collisions
cannot probe TMD PDFs and FFs directly as a function of $k_T$.  However, these
asymmetries do have sensitivity to the TMD PDFs and FFs integrated over $k_T$,
and attempts to describe the data phenomenologically using the Sivers and
Collins effects have been
done~\cite{Yuan:2008tv,Anselmino:2012rq,Anselmino:2013rya}.

Perturbative QCD calculations using collinear higher-twist quark-gluon 
correlations~\cite{Efremov:1981sh, Efremov:1984ip, Qiu:1991pp, Qiu:1998ia, 
kktwist30} can be performed and compared to data for inclusive SSAs in 
hadronic collisions.  While these correlation functions do not contain 
direct information on the transverse momentum distributions of partons, 
this approach has been related to $k_T$-moments of TMD PDFs and FFs such 
as the Sivers and Collins functions for multiparton correlations in the 
initial and final state, respectively~\cite{Ji:2006ub}.  Prior RHIC 
transverse SSA measurements for inclusive hadron production have been 
described relatively well by a combination of twist-3 effects in the 
initial and final states~\cite{kktwist31, kktwist32, Metz:2012ct, Kanazawa:2014dca}, 
but further refinement in both the theoretical calculations, for example 
through a better understanding of uncertainties, and in experimental 
measurements, for example through multi-differential measurements in more 
than one kinematic variable simultaneously, will be needed to test and 
understand these correlations in detail.

It has been predicted that TMD-factorization may be broken when the partonic
transverse momentum is explicitly taken into account, and the partons in the
two incoming protons can no longer be described by independent PDFs but instead
become correlated across the two protons~\cite{Rogers:2010dm}.  In this case
any phenomenology used to describe the asymmetries might become more complex,
depending on the size of the effects from factorization breaking.  The
breakdown of TMD-factorization leads to the prediction of additional spin
asymmetries in the case of hadron production in \ppup
collisions~\cite{Rogers:2013zha}, with the possible magnitude of any new
asymmetries still unknown.  These effects, due to color exchange, will be
interesting to explore further at RHIC once phenomenological predictions become
available.

This paper reports on measurements of the cross section and transverse 
single spin asymmetry for $\eta$ mesons at forward pseudorapidity 
(3.0\,$<$\,$|\eta|$\,$<$\,3.8) from the 2008 RHIC data taking period at 
$\sqrt{s}$\,=\,200~GeV.  A total integrated luminosity of 
$\mathcal{L}$\,=\,6.65~pb$^{-1}$ was sampled for these results.  
The measurement of different produced particle species will help to 
advance our understanding of the transverse SSAs ($A_N$) observed in \ppup 
collisions.  The comparison of pions, $\eta$ mesons, and kaons can shed 
light on initial- versus final-state spin-momentum correlations as well as 
possible isospin, strangeness, and mass effects.

A review of the RHIC polarized $p$$+$$p$ collider facility and the PHENIX 
experiment and detectors used for the measurements is given (Section 
\ref{sec:exp}), followed by a description of the analysis procedure 
(Section \ref{sec:yields}) used to procure the measurements of the cross 
section (Section \ref{sec:cross}) and transverse single spin asymmetry 
(Section \ref{sec:an}).  A final section is reserved for discussion of the 
results derived from these measurements.

\section{Experiment \label{sec:exp}}

\subsection{RHIC polarized $p$+$p$ collider}

\begin{figure}[thb]
  \centering
  \includegraphics[width=1.0\linewidth]{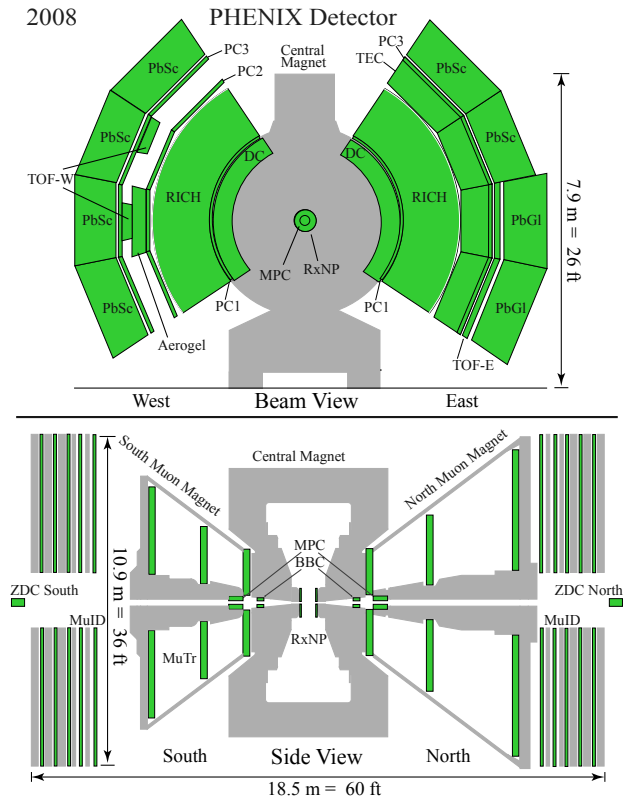}
  \caption{\label{fig:PHENIX_Detector} The PHENIX detector
    configuration during the 2008 RHIC run.}
\end{figure}

The Relativistic Heavy Ion Collider (RHIC) is a particle accelerator located at
Brookhaven National Laboratory.  RHIC has the capability of bunching, storing,
accelerating, and colliding polarized protons~\cite{rhic_polpp}, as well as
other ions, over a broad range of center-of-mass energies
($\sqrt{s}$\,=\,62.4~to~510\,GeV for polarized protons).  The injected beam
into RHIC is typically made up of 111 bunches of polarized protons, which
contain up to $\mathcal{O}(10^{11})$ protons per bunch for $p$+$p$ collisions
and are collided at several different points around the ring.  One such
interaction point is located at the PHENIX experiment \cite{phenix_overview}.
For the 2008 RHIC $p$$+$$p$ running, PHENIX (Fig.~\ref{fig:PHENIX_Detector})
consisted of two Spectrometer arms at central pseudorapidity
$|\eta|$\,$<$\,0.35, two Muon arms at pseudorapidity
1.2\,$<$\,$|\eta|$\,$<$\,2.4, two global detectors, and two calorimeters
(called the MPC detector) at forward pseudorapidity
3.1\,$<$\,$|\eta|$\,$<$\,3.9.

A key aspect of the asymmetry measurements is the ability to align the 
spin vectors of the protons in the beam in a desired direction. The 
net fraction of protons in the beam with their spin vectors aligned along 
this desired direction is called the polarization ($P$).  This must be 
measured to provide the correct scale for any asymmetry measurement. The 
polarization of the beams in RHIC is determined to within an uncertainty 
$\Delta P / P \sim 4\%$ using two different kinds of polarimeters: a 
Proton-Carbon polarimeter~\cite{Nakagawa.AIP2008} and a Hydrogen-Jet 
polarimeter~\cite{hjet_hard}.  The Proton-Carbon polarimeter provides fast 
relative measurements of the polarization several times during a fill, 
while the Hydrogen-Jet polarimeter measurement takes several hours but 
yields the absolute polarization.

The polarization direction alternates for consecutive bunches which 
minimizes potential time-dependent and spin-dependent systematic 
uncertainties.  In particular, detector efficiency and acceptance effects 
are minimized, as spin direction alternation in bunches allows use of the 
same detector for both polarization directions. During the 2008 RHIC run, 
the average clockwise beam (also known as the blue-beam) polarization was 
measured to be $P$\,=\,0.490\,$\pm$\,0.017, while the average 
counter-clockwise beam (yellow) polarization was 
$P$\,=\,0.410\,$\pm$\,0.015. The stable polarization direction in RHIC is 
transverse, \textit{i.e.}, perpendicular to the accelerator plane.

\subsection{PHENIX Local Polarimetry}

The polarization direction is also measured locally at PHENIX using a pair of
Zero-Degree Calorimeters (ZDCs). The ZDCs comprise two hadronic calorimeters,
located $\pm$18\,m from the nominal PHENIX interaction point. A shower maximum
detector (SMD) combined with the ZDC measures the transverse single spin
asymmetry of very forward ($\eta$\,$\gtrsim$6) neutrons which is found to be
nonzero, and as large as $A_N$\,$\sim$10\%~\cite{neutron_an0,neutron_an1}.  A
study of neutron $A_N$ in 2008 using the ZDC/SMD showed that the North-going
(blue) polarization axis was oriented off-vertical by
$\phi_{blue}$\,=\,0.263\,$\pm$\,0.03\,(stat)\,$\pm$\,0.090\,(syst)\,radians.
The South-going polarization axis was found to be consistent with the nominal
vertical direction,
$\phi_{yellow}$\,=\,0.019\,$\pm$\,0.048\,(stat)\,$\pm$\,0.103\,(syst).

\subsection{PHENIX Beam-Beam Counters}

The Beam-Beam Counters (BBC), see Fig.~\ref{fig:PHENIX_Detector}, comprise 
two arrays of 64 quartz \v{C}erenkov radiators connected to 
photomultiplier tubes (PMTs). The BBC is $z$\,=\,$\pm$144\,cm from the 
nominal interaction point and covers 3.0\,$<$\,$|\eta|$\,$<$\,3.9. The 
primary functions of this detector are to measure the position 
of the collision along the beam ($z$) axis to a precision of 
$\sigma(z_{vertex})$\,=\,2\,cm, to provide a minimally biased trigger, and 
to measure the luminosity.

\begin{figure}[thb]
\includegraphics[width=1.0\linewidth]{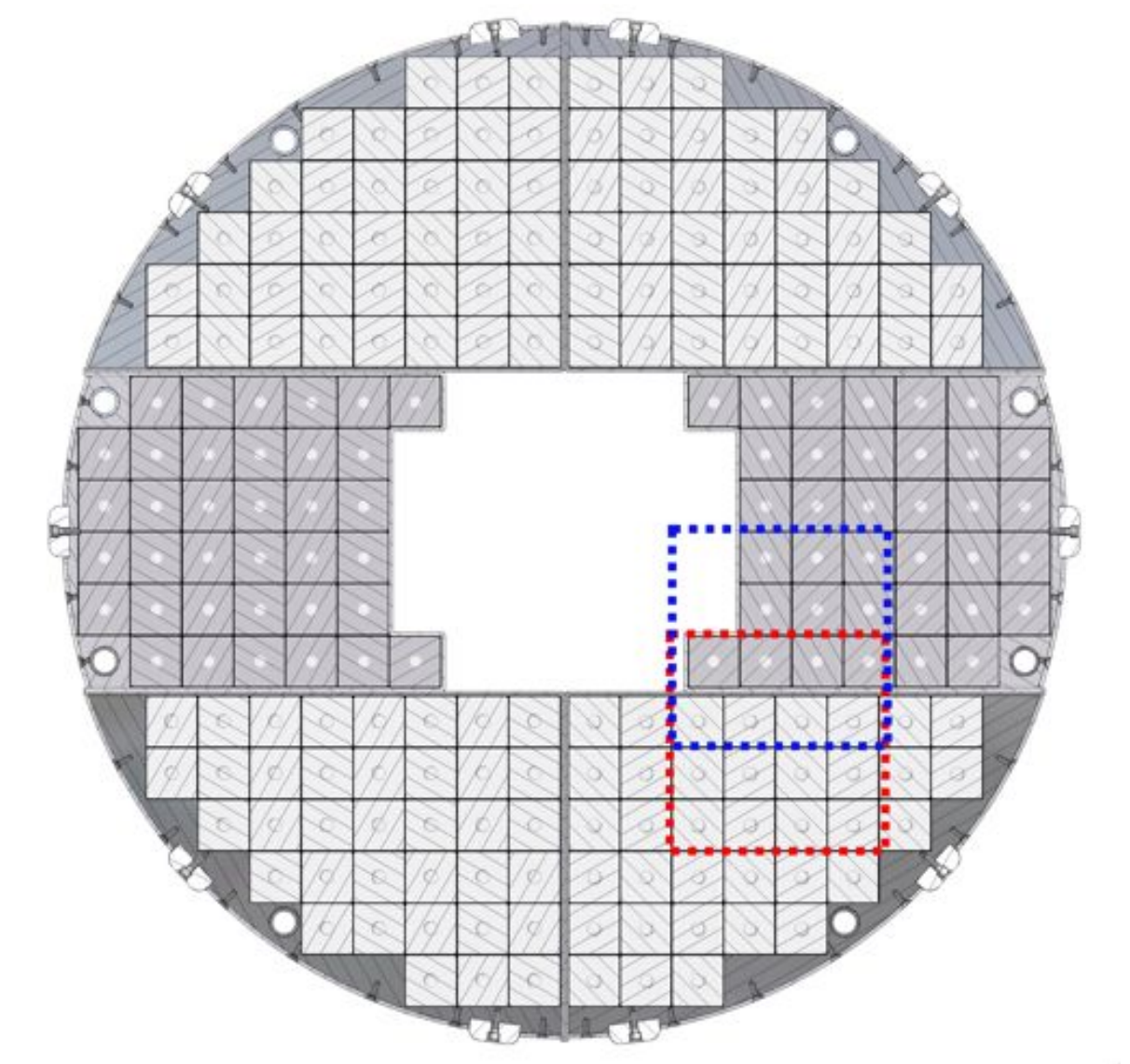}
\caption{\label{fig:mpc} (Color Online) A schematic of the North MPC as it appears in PHENIX.  The red and blue squares drawn on the MPC demonstrate an example of two overlapping 4$\times$4 trigger tiles.}
\end{figure}

\begin{figure*}[thb]
  \includegraphics[width=1.0\linewidth]{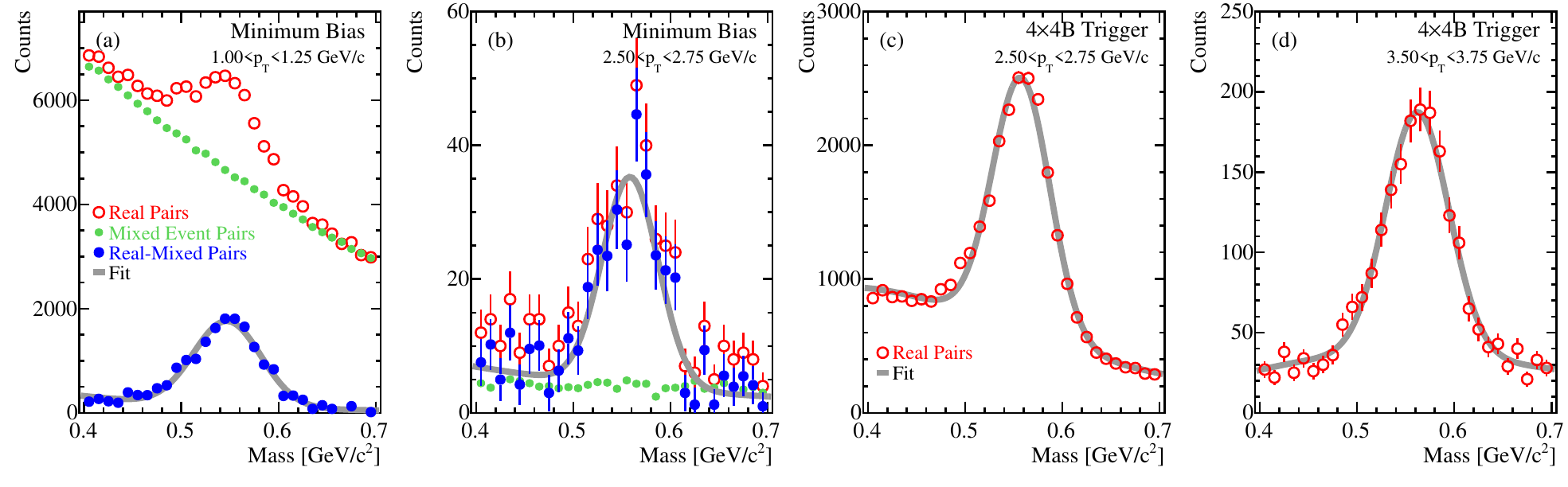}
  \caption{\label{fig:mixedevent} (Color online) The invariant mass distribution
    for minimum bias (panels (a) and (b)) and \highpt (panels (c) and (d))
    samples.  In all panels, open red circles represent all real pairs formed
    from MPC clusters.  In panels (a) and (b) the small green closed symbols
    show the combinatorial background from mixed events (see text) and the
    closed blue symbols show the combinatorial-subtracted real pairs.  Panels
    (b) and (c) show the same $p_{T}$ selection and illustrate the importance
    of triggering to enhance the statistical significance at large momenta.  
    Grey lines show the fit to the data used to extract the yield.}
\end{figure*}

\subsection{PHENIX MPC detector}

The Muon Piston Calorimeter (MPC) comprises two forward electromagnetic 
calorimeters, referred to as the South and North MPC, see 
Fig.~\ref{fig:PHENIX_Detector}, placed $\pm$220\,cm from the nominal 
interaction point along the beam axis. The South (North) MPC is made up of 
196 (220) $2.2 \times 2.2 \times 18$\,cm$^{3}$ $PbWO_{4}$ crystal towers,
and is read out with Hamamatsu S8664-55 avalanche photo-diodes (APD). The
MPC covers the pseudorapidity regions -3.7~$<$~$\eta$~$<$~-3.1 and 
3.1\,$<$\,$\eta$\,$<$\,3.9, respectively.  The primary goal of the MPC is 
to identify $\pi^{0}$ and $\eta$ mesons and measure their energy.

$PbWO_{4}$ crystals were chosen for their short radiation length (0.89\,cm) and
small Moli\`{e}re radius (2.0\,cm).  Similar $PbWO_4$ crystals were originally
used and extensively tested for the PHOS detector~\cite{Ippolitov:2005wi}, part
of the ALICE experiment at CERN.  The MPC is not cooled, and runs at the
ambient temperature of its location in PHENIX.  The gain variation with time,
due largely to temperature variations and radiation damage to the crystals and
APDs, is tracked using a LED calibration system.  The absolute gain calibration
comprises the LED tracking and tower by tower calibrations using $\pi^0$s.
The relative energy resolution after calibration was found to be $\sigma(E)/E =
13\%/\sqrt{E} \oplus 8\%$.  Comparisons between the $\pi^0$ and
$\eta$ meson using real data and simulations showed that an overall energy
scale uncertainty of 2\% remained after all the calibrations, and also
determined that the position resolution for clusters was about 2 mm. A
schematic of the North MPC is given in Fig.~\ref{fig:mpc}.

\subsection{Triggers}

Readout of the PHENIX detector was done using one of two independent triggers
for this analysis.  The \minbias (MB) trigger initiated readout when at least
one BBC PMT in each array is hit, and when the collision vertex is within
$|z|$\,$<$\,30\,cm of the nominal interaction point in PHENIX.  As the number
of collisions delivered by RHIC exceeds the data-taking rate of the PHENIX data
acquisition system, only a small fraction of events can be recorded with
``minimum bias.'' To enhance the more rare (higher momentum) $\eta$ mesons in
the data stream an additional trigger is used to record the high-$p_T$ part of
the cross section.  This higher momentum trigger (called the \highpt trigger)
records an event when the total sum in any of the 4$\times$4 trigger arrays of
MPC towers satisfies an energy threshold of $E\gtrsim20$\,GeV.  The 4$\times$4
trigger arrays are particular groupings of towers and are called
\textit{tiles}.  Each tile overlaps by 2 towers in the horizontal and vertical
directions, as shown in Fig.~\ref{fig:mpc}, to provide even coverage for the
trigger over the whole detector.  The \highpt trigger is formed without the
requirement of a collision vertex from the BBCs.

\section{Identification of $\eta$ mesons in the MPC \label{sec:yields}}

To identify $\eta$ mesons in the MPC, the decay channel 
$\eta\rightarrow\gamma\gamma$ is used which has a branching ratio of 
$BR$\,=\,0.3941\,$\pm$\,0.0020~\cite{pdg0}. Clusters of MPC towers from a 
single event are combined to form photon candidates.  To increase the 
likelihood that a cluster is due to a real photon, clusters which do not 
possess the characteristic electromagnetic shower shape are discarded.  
Clusters with their central tower tagged as noisy or inactive are also 
removed from the analysis.  Once a sample of clusters is reduced to an 
enhanced sample of real photon candidates, clusters are paired together 
and an invariant mass is calculated, Eq.~\ref{eq:mgg_equ}:

\begin{equation}
\displaystyle M_{\gamma \gamma} = \sqrt{4 \cdot E_{\rm 1} \cdot E_{\rm 2}} \cdot \sin(\theta_{12} / 2)
\label{eq:mgg_equ}
\end{equation}

\noindent where $E_{\rm 1,\rm 2}$ is the measured energy of each 
cluster, and $\theta_{12}$ is the opening angle between the momentum 
vectors of the two clusters.  Additional kinematic cuts are made on paired 
clusters for the \minbias and \highpt data sets.  A minimum energy $E_{\rm 
1}+E_{\rm 2}$\,$>$\,7\,GeV and 10\,GeV respectively is imposed. A 
maximum energy asymmetry, $\alpha$\,=\,$|\frac{E_{\rm 1} - E_{\rm 
2}}{E_{\rm 1} + E_{\rm 2}}|$, of $\alpha$\,=\,0.6~and~0.8, 
respectively, is required. The difference in the energy asymmetry cut 
between the two triggers is due to differences in the signal to background 
figure of merit.  Finally, the separation between the two clusters
$\Delta R$ has to be greater than 2.6~cm, minimizing merging effects between cluster 
showers. After the application of these cuts, the invariant mass 
is calculated for all pairs, which is shown in Fig.~\ref{fig:mixedevent} 
as open symbols.

\section{The $\eta$ Meson Cross Section \label{sec:cross}}

The cross section can be written in terms of measured quantities as:

\begin{equation}
\displaystyle E \frac{d^{3}\sigma_{h}}{dp^{3}} = \frac{1}{\mathcal{L}_{\rm pp,inel}} \frac{1}{2\pi p_{T}} \frac{\Delta N^{\rm meas}_{\eta}}{BR \cdot \epsilon_{\rm reco} \cdot \epsilon_{\rm trig}\Delta p_{T} \Delta y}
\label{eq:equ_cross}
\end{equation}

\noindent where $\Delta N^{\rm meas}_{\eta}$ is the number of measured (raw)
$\eta$ mesons over a rapidity range $\Delta y$ and transverse momentum interval
$\Delta p_{T}$.  Note $\Delta y$\,$\approx$\,$\Delta \eta$ for $\eta$ mesons at
forward rapidity at the $p_T$ measured in this analysis.  The data are scaled
by the integrated luminosity ($\mathcal{L}_{\rm pp,inel}$) and the branching
fraction, $BR$, for this decay channel. To account for inefficiency in
triggering and reconstruction, the $\Delta N^{\rm meas}_{\eta}$ is corrected by
factors $\epsilon_{\rm trig}$ and $\epsilon_{\rm reco}$ respectively.  Each of
these components is described in the following sections.

\subsection{Integrated Luminosity ($\mathcal{L}_{\rm pp,inel}$)}

The luminosity is calculated as the ratio of the number of minimum bias 
events sampled for each trigger condition, within $|z|<30$cm, divided by the 
part of the $p$$+$$p$ cross section to which the BBCs are sensitive.  This cross 
section is $\sigma_{\rm pp}^{\rm BBC}$\,=\,23.0\,$\pm$\,2.3\,mb which 
is determined using a Vernier Scan procedure~\cite{vernier}. The total 
integrated luminosity of the \minbias dataset is $\mathcal{L}_{\rm 
MB}$\,=\,0.0192\,pb$^{-1}$ and that of the \highpt dataset is 
$\mathcal{L}_{\rm 4\times 4B}$\,=\,3.87\,pb$^{-1}$.

\subsection{Yield extraction ($\Delta N^{\rm meas}_{\eta}$)}

The invariant mass distribution (Fig.~\ref{fig:mixedevent}) has two 
distinct components: correlated pairs (for example from $\eta$ meson 
decays) and uncorrelated (combinatorial) background pairs, due to pairing 
of clusters from different parent sources. To account for this 
combinatorial background in the \minbias dataset 
(0.5\,$<$\,$p_T$\,$<$\,3.0\,GeV/$c$), photon candidates are analyzed from 
different events (which necessarily removes all real combinations) to form 
a mixed event distribution. The mixed event pair distribution is 
normalized (green closed circles in Fig.~\ref{fig:mixedevent}) to the real 
pair distribution by taking the ratio of the real and mixed distributions 
and fitting with a constant at high invariant mass, and then subsequently 
scaling the mixed event distribution by this constant.  The subtraction 
from this real pair distribution results in a final $\gamma\gamma$ 
invariant mass spectrum which has all uncorrelated background pairs 
removed (blue closed circles in Fig.~\ref{fig:mixedevent}). Using the same 
mixed event procedure, only a small fraction of the \highpt background was 
found to be uncorrelated, the rest is made from a jet correlated 
background made primarily from $\pi^{0}$ decays.  The mixed-event 
subtraction removes only a small fraction of the uncorrelated background 
in the \highpt triggered dataset, so it is not applied in this case (see 
panels (c) and (d) in Fig.~\ref{fig:mixedevent}).

Raw yields are extracted by fitting the invariant mass distributions 
(mixed-event subtracted in \minbias sample) with a function for the correlated 
background plus a constant term times a normalized Gaussian distribution 
representing the signal peak (gray lines in Fig. \ref{fig:mixedevent}).  
The optimal background function for the \minbias (\highpt) dataset was an 
exponential (Gamma Distribution) function.  Variation of the functional 
form of the background (2nd, 3rd order polynomial) was used to evaluate 
the systematic uncertainty on the yield extraction.

\subsection{Efficiency Corrections ($\epsilon_{\rm reco}$ and 
$\epsilon_{\rm trig}$)}

Measured (raw) yields must be corrected for reconstruction and trigger 
inefficiencies.  Simulations are used to calculate the reconstruction 
efficiency ($\epsilon_{\rm reco}$), which corrects for geometric 
acceptance and detector resolution effects.  To produce an $\eta$ meson 
$p_T$ spectrum which is similar to that in real data, a full Monte Carlo 
sample of single $\eta$ mesons are initially generated flat in $p_T$ and 
pseudorapidity in the MPC kinematics, and with the same $z$-vertex 
distribution as measured in data. These generated single $\eta$ mesons are 
passed through a {\sc geant} (3.21)~\cite{geant0} description of the 
PHENIX detector and subsequent energy deposits are embedded into real data 
\minbias events.  \Minbias events here do not necessarily contain an 
$\eta$ meson from the collision.  The same cluster identification and pair 
cuts are applied, followed by the full reconstruction, similar to that in 
the real data analysis.

\begin{figure}[thb]
  \includegraphics[width=1.0\linewidth]{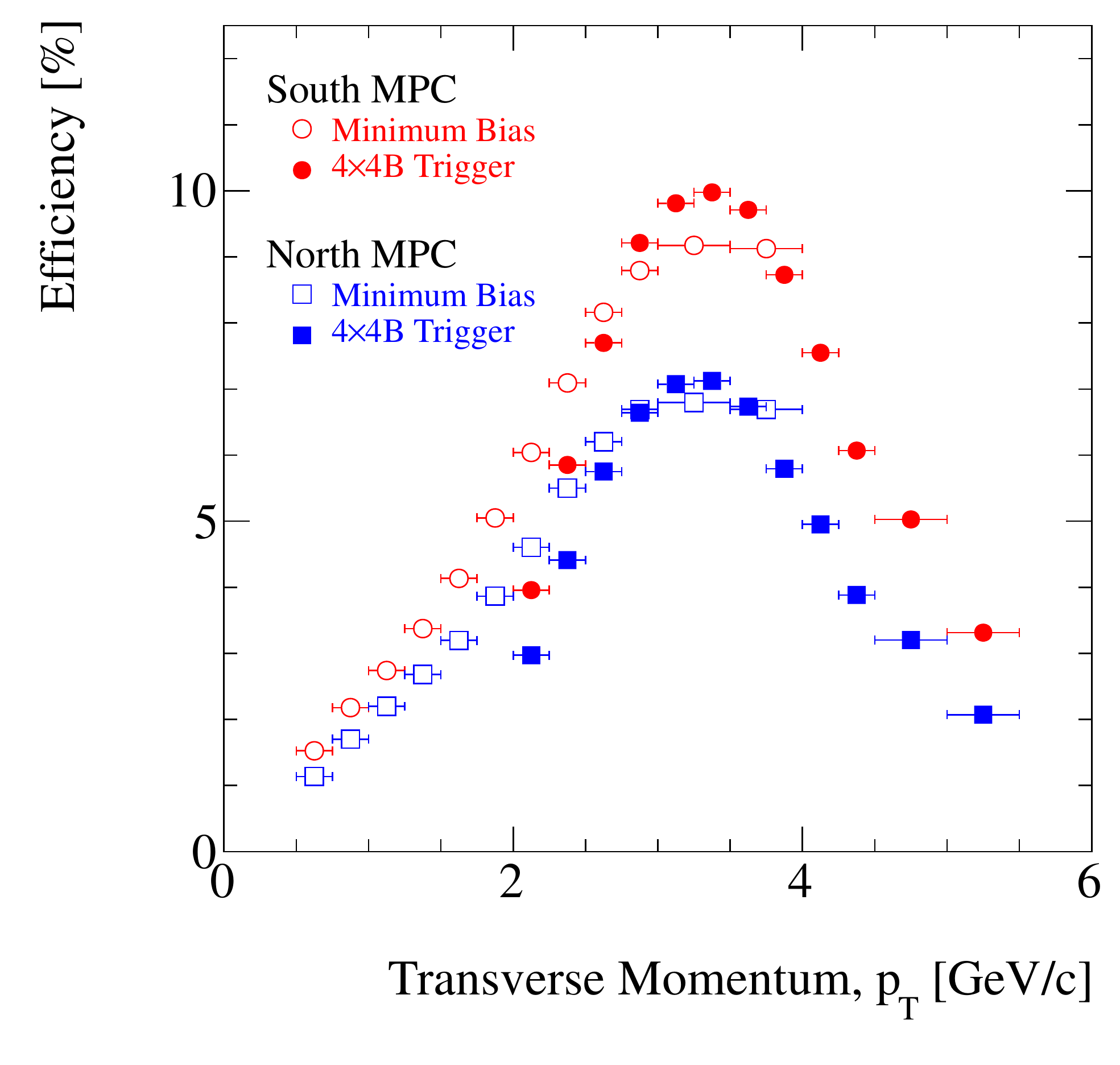}
  \caption{\label{fig:effreco} (Color online) Reconstruction efficiency for
    $\eta$ mesons using the \minbias (\highpt) dataset, shown as open (closed) symbols.
    The red/circle (blue/square) symbols show the $\eta$ meson reconstruction
    efficiency for the South (North) MPC.}
\end{figure}

The next step weights the reconstructed and generated $\eta$ mesons in 
$p_T$ and pseudorapidity to mimic the measured data distribution. This 
accounts for $p_T$ smearing effects on an exponential spectrum, and for 
the falling pseudorapidity dependence in the forward region. As the 
weighting is dependent on the shape of the corrected spectrum, an 
iterative procedure is used to ensure the efficiency correction converges 
to a stable value. The reconstruction efficiency is calculated as the 
ratio of reconstructed $\eta$ mesons divided by the number generated. The 
reconstruction efficiency for the South and North MPC for both triggers is 
shown in Fig.~\ref{fig:effreco}.  The North MPC has a lower reconstruction 
efficiency than the South, due to a more restrictive noisy/inactive tower 
map in the North.  The reconstruction efficiency shape is predominantly 
due to the geometric acceptance coupled to the narrowing $\gamma\gamma$ 
opening angle from low to higher momenta.  At low momenta, wider opening 
angles can prohibit the measurement of both $\gamma$s in the detector.  
At high momenta, cluster merging increasingly inhibits the detection of 
distinct $\gamma$ pairs. Significant cluster merging effects occur when 
the cluster separation is less than 1.5 times the tower width ($\Delta 
R$\,$<$\,3.3~cm).

The trigger efficiency ($\epsilon_{\rm trig}$) is estimated by taking the 
ratio of $\eta$ meson yields found using the trigger of interest (for 
example \minbias) in coincidence with any other trigger which is unrelated 
(unbiased) divided by the same unrelated trigger without the coincidence 
requirement,

\begin{equation}
\label{eq:trigeff}
\displaystyle \epsilon_{\rm trig}^{\eta} = \frac{N^{\eta}_{\rm unbias \wedge trig} }{N^{\eta}_{\rm unbias}}
\end{equation}

For the \minbias trigger efficiency, $\epsilon^{\eta}_{\rm MB}$, the 
\highpt trigger is used as this maximizes the $\eta$ meson yield 
statistics.  The measured \minbias trigger efficiency is found to be 
$\epsilon^{\eta}_{\rm MB}$\,=\,0.76 \,$\pm$\,0.01 
(stat)\,$\pm$\,0.06\,(syst). There is a slight dependence on $p_T$, which 
has been factored into the systematic uncertainty.
  
\begin{figure*}[thb]
  \includegraphics[width=1.0\linewidth]{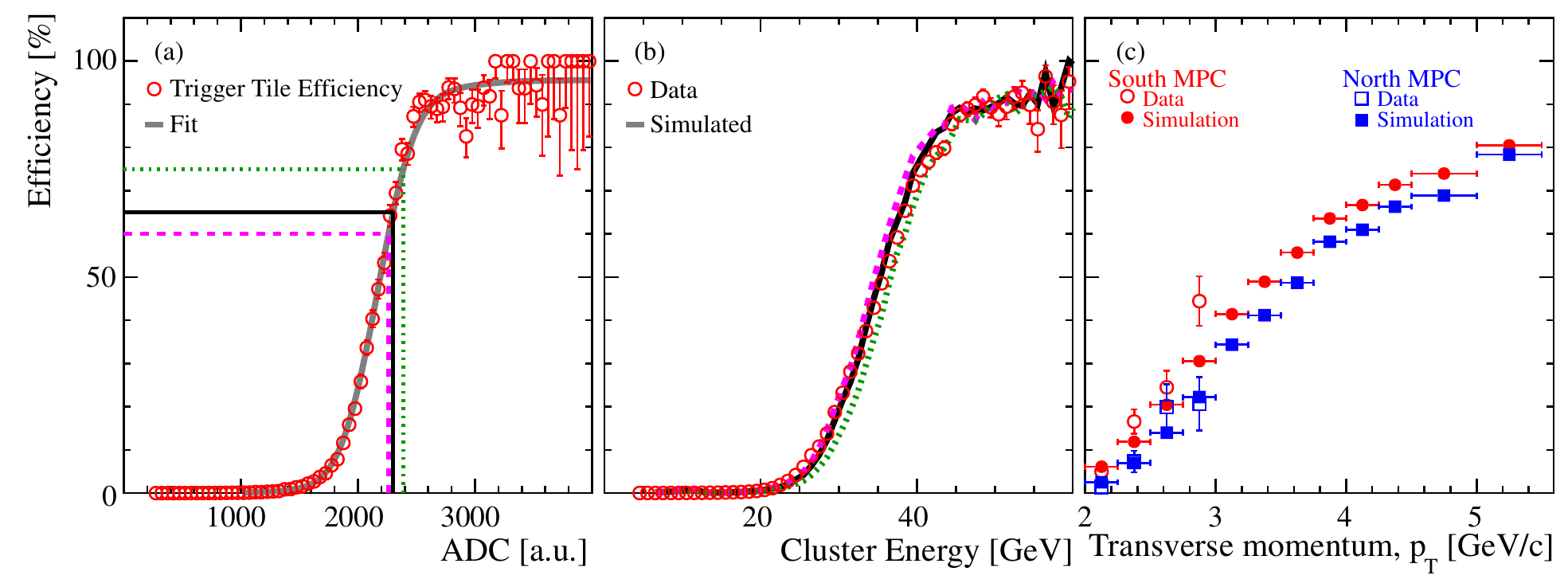}
  \caption{\label{fig:trigeff}  (Color online) Panel (a) shows the
    trigger efficiency for a single \tile tower array in the \highpt trigger.  The solid, dashed, and dotted lines represent $\theta_{\rm thresh}$\,=\,0.66,\,0.60,\,and\,0.75, respectively.  Panel (b)
    shows a comparison of the cluster efficiency as a function of energy to
    the simulated efficiency generated using the different $\theta_{thresh}$.  Panel (c)
    shows the $\eta$ meson \highpt trigger efficiency,
    $\epsilon^{\eta}_{\rm 4\times4B}$ (systematic error not included).  The open symbols represent
    $\epsilon^{\eta}_{\rm 4\times4B}$ calculated using Eq.~\ref{eq:trigeff}
    with the \minbias trigger as the unrelated trigger.
    The closed points represent $\epsilon^{\eta}_{\rm 4\times4B}$ calculated
    from simulation.  South (North) efficiencies are shown
    as circles (squares).}
\end{figure*}

For the \highpt trigger efficiency for $\eta$ mesons, 
$\epsilon^{\eta}_{\rm 4\times 4B}$, the \minbias trigger is used as the 
unrelated trigger.  The statistics in the \minbias sample is limited, 
however, and the efficiency can only be determined from the data up to 
$p_{T}$\,$<$\,3.0~GeV/$c$ (see panel (c) in Fig.~\ref{fig:trigeff}, open 
symbols). Instead, the trigger efficiency for $\eta$ mesons in the \highpt 
triggered sample is calculated by simulating the \highpt trigger.

The \highpt trigger comprises a total of 56 (61) overlapping \tile tower 
array sums from the South (North) MPC.  An example of the efficiency of an 
individual \tile array from data is shown in panel (a) in 
Fig.~\ref{fig:trigeff}. This efficiency is fit with a double error 
function

\begin{equation}
\displaystyle f(x) = \int_{-\infty}^{x} [ a g_1({x}') + (1 - a) g_2{({x}')}]d{x}'
\label{eq:errfunc}
\end{equation}

\noindent where $g_1(x)$ and $g_2(x)$ are Gaussian distributions. The 
efficiency curve shown in panel (a) of Fig.~\ref{fig:trigeff} covers the 
entire data-taking period, and relative gain changes throughout the RHIC 
run due to temperature variations and radiation damage to the detector 
cause a large spread in the rise of the efficiency curve.  This gain 
variation is monitored with an LED calibration system.  The trigger 
threshold ($\theta_{\rm thresh}$) at any given instant is a step function, 
and is thus implemented in the simulation as a step-function.  The changes 
in the effective threshold due to the gain variation over the run are 
accounted for in the simulation by varying the threshold using the data 
from the LED monitoring.  Fit parameters from the 117 different trigger 
tile efficiency curves are derived and used in the trigger simulation to 
determine an optimal $\theta_{\rm thresh}$, and thus trigger efficiency 
for $\eta$ mesons.

To tune the trigger simulation, reconstructed $p$+$p$ events from {\sc pythia}
(tune-A)~\cite{PythiaManual:2006} were processed through the trigger simulation
and matched to real data.  The cluster trigger efficiency is well reproduced in
the simulation when using a mean tile trigger threshold of $\theta_{\rm
thresh}$\,=\,0.66, determined from the fit parameters in Eq.~\ref{eq:errfunc}
(see solid line in panels (a) and (b)).  On average, $\theta_{\rm thresh}$
corresponds to $\langle E_{\rm 4\times4} \rangle$\,$\approx$\,40\,GeV. The
comparison is shown in panel (b) of Fig.~\ref{fig:trigeff}, where good
agreement is seen between the simulation of the \highpt trigger and the data
efficiency curve to all energies of interest in this analysis. Variations of
the threshold in the simulation between 0.60\,$<$\,$\theta_{\rm
thresh}$\,$<$\,0.75 (see dotted (0.60) and dashed (0.75) lines in panels (a)
and (b)) are used to estimate a systematic uncertainty on reproducing the
\highpt cluster trigger efficiency.  These systematic variations account for
differences in the South and North MPC, and for a turn-on uncertainty which
occurs for low energy clusters that are smeared out above and below the
selected trigger turn-on.

Within this trigger simulation framework, the \highpt trigger 
efficiency for $\eta$ mesons is calculated from the same single-$\eta$ 
simulations used in the reconstruction efficiency study.  This simulation 
accounts for effects such as when the distance between the two decay 
photons, $\Delta R$, is small enough that the two photons fall into the 
same \tile tile such that their energy sum fires the trigger together.  
Panel (c) in Fig.~\ref{fig:trigeff} shows the $\eta$ meson \highpt trigger 
efficiency calculated via simulation, with a comparison to the 
statistically limited values measured from the \minbias trigger in the 
overlap region of 2.0\,$<$\,$p_{T}$\,$<$\,3.0\,GeV/$c$ calculated using 
Eq.~\ref{eq:trigeff}.  In this overlap region there is good agreement within 
statistical (shown) and systematic uncertainties (not shown, see next section).

\subsection{Systematic Uncertainties}

The systematic uncertainties are divided into three types. Statistical and 
point-to-point uncorrelated systematic uncertainties are added in 
quadrature to form type-A uncertainties.  Type-B represent correlated 
uncertainties between \pT bins.  Type-C are external global systematic 
uncertainties which underlay the measurement.

The functional form of the background used in the yield extraction was 
varied and contributes 5-15\% to the type-A uncertainties. The 
systematic uncertainty due to energy scale (type-B) was found to vary from 
3 to 30\% for $p_T$\,=\,0.5 to 5.0\,GeV/$c$. A global reconstruction 
efficiency uncertainty (type-B) of 11.5\,\% (27.5\%) is applied for 
$p_T$\,$>$\,0.75 GeV/$c$ ($p_T$\,$<$\,0.75 GeV/$c$).  An additional 
reconstruction efficiency uncertainty of 1 to 20\% for $p_T$\,=\,3.0 to 
5.0\,GeV/$c$ is assigned due to cluster merging effects (type-B).  The 
systematic uncertainty on varying the turn-on threshold (type-B) for the 
\highpt trigger efficiency leads to 30\% uncertainty at 
$p_T$\,=\,2.0\,GeV/$c$, which decreases exponentially to 5\% at 
$p_T$\,=\,5.0 GeV/$c$. A further global (type-C) systematic uncertainty of 
9.7\,\% is applied based on the luminosity monitoring of the BBC.

\subsection{Cross Section Results}

The cross section is calculated using Eq.~\ref{eq:equ_cross} independently 
for the South and North MPC, and for both the \minbias and \highpt data 
sets.  For both trigger conditions, the South and North reconstructed 
cross section agree to within 2\% across $p_T$.  The South and North 
cross section measured for each trigger are weighted together to determine 
the final cross section spectrum.  Agreement in the overlap region 
(2\,$<$\,$p_T$\,$<$\,3\,GeV/$c$) between the \minbias and \highpt cross 
section was within 7\% across $p_T$ and are within the type-A systematic 
uncertainties.  In the overlapping region, data points from the two data 
sets are combined as a weighted average.

\begin{figure}[thb]
  \includegraphics[width=1.0\linewidth]{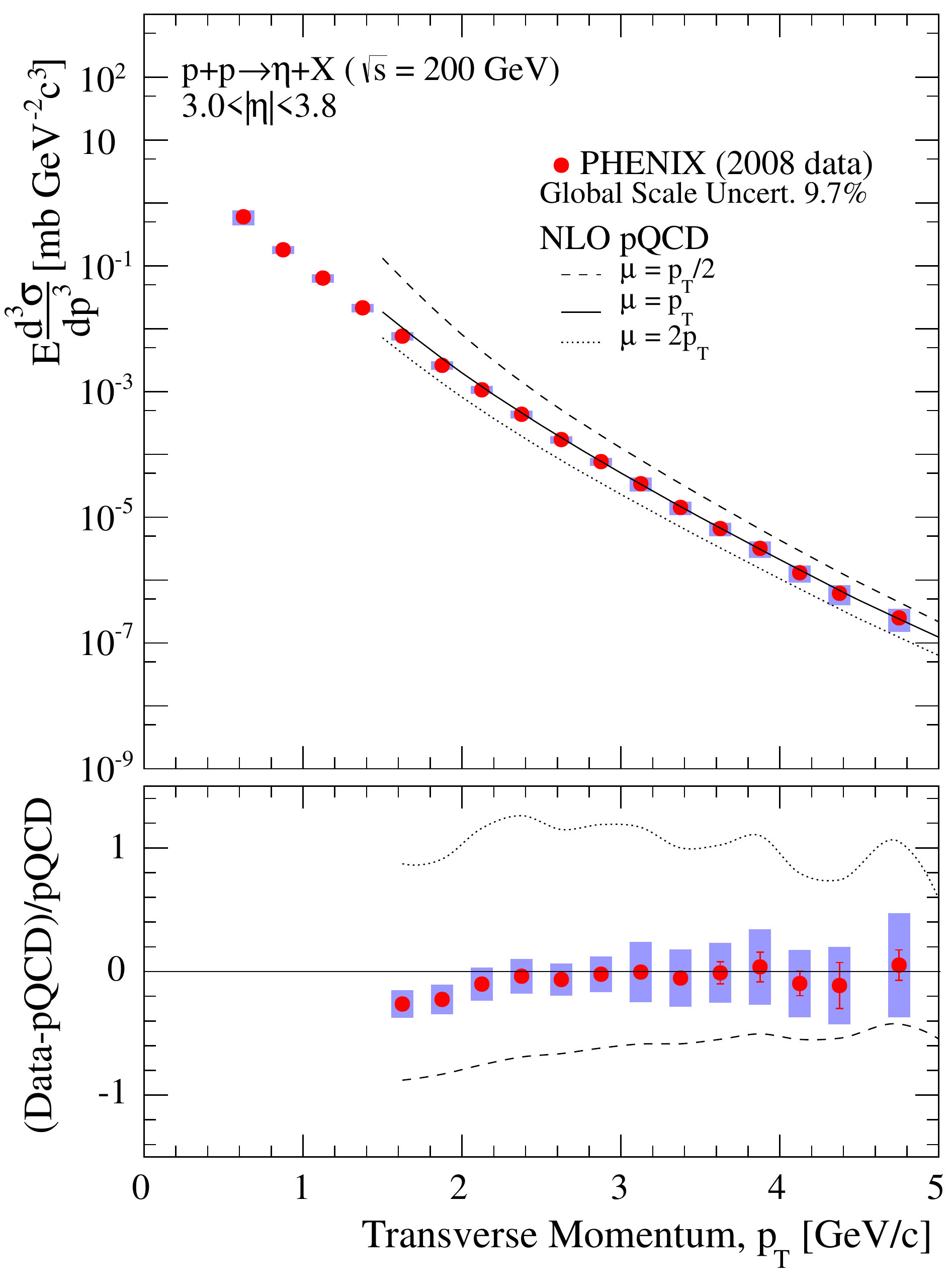}
  \caption{\label{fig:final_cross} (Color online) The cross section of inclusive
    $\eta$ mesons produced from $p$+$p$ collisions at
    $\sqrt{s}$\,=\,200\,GeV at forward rapidity.
    The upper panel shows the measured cross section versus transverse
    momentum ($p_T$), compared to an NLO pQCD
    calculation at three different scales $\mu$~\cite{marco_pqcd0, marco_pqcd1}.
    The lower panel shows the difference between the measured cross
    section and each of the NLO pQCD calculations.  Error bars (bands)
    represent type-A (type-B) systematic uncertainties.  A global
    scale uncertainty (type-C, 9.7\%) is due to the luminosity and global reconstruction
    uncertainties. }
\end{figure}

The invariant cross section of $\eta$ mesons is shown in 
Figure~\ref{fig:final_cross} and Table~\ref{tab:cross} as a function of 
transverse momentum, measured between 0.5\,$<$\,$p_T$\,$<$\,5.0~GeV/$c$ 
within a pseudorapidity range of 3.0\,$<$\,$|\eta|$\,$<$\,3.8.  The results 
are compared to an NLO pQCD calculation for three different choices of scale $\mu$~\cite{marco_pqcd0, marco_pqcd1}, over the same pseudorapidity region as the measurement.  Here, $\mu$ represents the factorization, renormalization, 
and fragmentation scales, which are set to be equal to one another.
   
The lower panel shows the comparison between the measured cross section 
and the NLO pQCD. For $p_T$\,$>$\,2.0\,GeV/$c$, the NLO pQCD calculation 
is in very good agreement with the measured cross section. Upon 
approaching the pQCD limit at low momentum 
($p_T$\,$<$\,2.0\,GeV/$c$) the agreement is less clear, but well within 
the factorization uncertainty.

\begin{table}
\caption{\label{tab:cross} The measured $\eta$ meson cross section versus 
$p_T$ at forward rapidity for the 2008 dataset with statistical and 
systematic (type-A and type-B) uncertainties.  There is an additional 
normalization uncertainty of 9.7\% (type-C).}
\begin{center}
\begin{ruledtabular} \begin{tabular}{cccc}

{$p_{T}$}  & {$E \frac{d^3 \sigma}{d p^3}$} & {Type-A} & {Type-B}   \\
$[{\rm GeV/}c]$ & {[mb GeV$^{-2}c^{3}$]} \\
\hline
0.625 & 6.03$\times$10$^{-1}$ & 8.76$\times$10$^{-2}$ & 1.68$\times$10$^{-1}$ 
\\ 
0.875 & 1.80$\times$10$^{-1}$ & 3.12$\times$10$^{-2}$ & 2.61$\times$10$^{-2}$
\\ 
1.125 & 6.39$\times$10$^{-2}$ & 4.48$\times$10$^{-3}$ & 9.71$\times$10$^{-3}$
\\ 
1.375 & 2.15$\times$10$^{-2}$ & 8.17$\times$10$^{-4}$ & 3.35$\times$10$^{-3}$
\\ 
1.625 & 7.61$\times$10$^{-3}$ & 3.98$\times$10$^{-4}$ & 1.17$\times$10$^{-3}$
\\ 
1.875 & 2.61$\times$10$^{-3}$ & 1.31$\times$10$^{-4}$ & 4.08$\times$10$^{-4}$
\\ 
2.125 & 1.07$\times$10$^{-3}$ & 5.31$\times$10$^{-5}$ & 1.59$\times$10$^{-4}$
\\ 
2.375 & 4.35$\times$10$^{-4}$ & 2.04$\times$10$^{-5}$ & 6.33$\times$10$^{-5}$
\\ 
2.625 & 1.72$\times$10$^{-4}$ & 6.39$\times$10$^{-6}$ & 2.39$\times$10$^{-5}$
\\ 
2.875 & 7.68$\times$10$^{-5}$ & 3.08$\times$10$^{-6}$ & 1.13$\times$10$^{-5}$
\\ 
3.125 & 3.42$\times$10$^{-5}$ & 1.19$\times$10$^{-6}$ & 8.42$\times$10$^{-6}$
\\ 
3.375 & 1.43$\times$10$^{-5}$ & 8.87$\times$10$^{-7}$ & 3.53$\times$10$^{-6}$
\\ 
3.625 & 6.61$\times$10$^{-6}$ & 5.96$\times$10$^{-7}$ & 1.62$\times$10$^{-6}$
\\ 
3.875 & 3.20$\times$10$^{-6}$ & 3.71$\times$10$^{-7}$ & 9.41$\times$10$^{-7}$
\\ 
4.125 & 1.31$\times$10$^{-6}$ & 1.42$\times$10$^{-7}$ & 3.95$\times$10$^{-7}$
\\ 
4.375 & 6.17$\times$10$^{-7}$ & 1.30$\times$10$^{-7}$ & 2.17$\times$10$^{-7}$
\\ 
4.750 & 2.51$\times$10$^{-7}$ & 2.92$\times$10$^{-8}$ & 1.01$\times$10$^{-7}$
\\ 
\end{tabular} \end{ruledtabular}
  \end{center}

\end{table}

\section{The Transverse Single Spin asymmetry for $\eta$ Mesons \label{sec:an}}

In polarized $p^{\uparrow}$+$p$ collisions, the cross section of hadron 
production can be modified in azimuth, with respect to the polarization 
direction. To first order the azimuthally dependent cross section can be 
written as

\begin{equation}
\displaystyle \dfrac{\rm d \sigma}{\rm d \Omega} = \left(\dfrac{\rm d \sigma}{\rm d \Omega}\right)_{0} \left(1 + P_y \cdot A_N \cdot \cos\phi \right)
\label{eq:sigma_an}
\end{equation}

\noindent where $\left(\dfrac{\rm d \sigma}{\rm d \Omega}\right)_{0}$ is 
the unpolarized differential cross section, $P_y$ is the vertical beam 
polarization, and $A_N$ is the transverse single spin asymmetry.  This 
dependence can be measured as

\begin{equation}
\displaystyle P_y \cdot A_N \cdot \cos\phi = \epsilon_N(\phi)
\label{eq:an_mod}
\end{equation}

\noindent where $\epsilon_N(\phi)$ is the measured raw asymmetry which, to 
first order, is an azimuthal cosine modulation.  For this analysis, $A_N$ 
is found by first measuring the raw asymmetry ($\epsilon_N(\phi)$), fitting 
it with a cosine function, and then dividing the amplitude by the average 
beam polarization. The raw asymmetry is measured in this analysis using 
two methods~\cite{measan0}.

The first method is known as the polarization formula,

\begin{equation}
\displaystyle \epsilon^{\rm pol}_N(\phi) = \frac{N^\uparrow(\phi) - N^\downarrow(\phi)}{N^\uparrow(\phi) + N^\downarrow(\phi)}
\label{eq:eps_pol}
\end{equation}

\noindent which uses two different polarization yields (up - $\uparrow$ 
and down - $\downarrow$) in one azimuthal region. This method is preferred 
if the acceptance is not homogeneous, but relative luminosity effects 
($\mathcal{R}$\,=\,$\frac{\mathcal{L}^{\uparrow}}{\mathcal{L}^{\downarrow}}$) 
must be taken into account.

A second method is known as the square-root formula, 
Eq.~\ref{eq:eps_sqrt}, which uses the geometric mean of the yields $N$ 
from two azimuthal regions on opposite sides of the MPC ($\phi$ and 
$\phi$+$\pi$) and two polarization directions (up -- $\uparrow$ and down 
-- $\downarrow$).  When there is little loss of acceptance, particularly 
dead areas in azimuthal space, this method is advantageous as detector 
efficiency and relative luminosity effects cancel.

\begin{equation}
\displaystyle \epsilon^{\sqrt{\phi}}_N = \frac{\sqrt{N ^\uparrow(\phi) \cdot 
N^\downarrow(\phi + \pi)} - \sqrt{N ^\downarrow(\phi) \cdot N^\uparrow(\phi + \pi)}}{\sqrt{N ^\uparrow(\phi) \cdot N^\downarrow(\phi + \pi)} + \sqrt{N ^\downarrow(\phi) \cdot N^\uparrow(\phi + \pi)}}
\label{eq:eps_sqrt}
\end{equation}

\noindent The final transverse single spin asymmetry result reported uses 
the square-root formula.  The polarization formula serves as a cross 
check.

\subsection{Polarization}

To measure $A_N$, the polarization and spin information of only one beam 
is used, while the other beam's spin information is ignored, such that it 
is integrated over to a net polarization of zero. As one chooses which 
beam to use as `polarized', two independent $A_N$ measurements can be 
made: one utilizing the North-going beam's polarization, and one utilizing 
the South-going beam's polarization.  Effectively, as the South and North 
MPC detectors are independent with differing systematics, two independent 
measures of $A_N$ are derived, allowing for more reliable evaluation of 
systematic uncertainties on the results.

\subsection{$A_N$ Analysis}

To measure the raw $A_N$, the $\phi$ distribution of the reconstructed 
$\eta$ meson is divided into twelve azimuthal bins, and spin dependent 
$\eta$ meson yields are obtained for each bin.

To extract the $\eta$ meson yields for the $A_N$ measurements, the 
invariant mass spectra from all photon pairs are first formed independent 
of spin direction and $\phi$, binned in $x_F$ (or $p_T$).  These invariant 
mass spectra are then fit with a signal Gaussian and background function. 
The signal Gaussian establishes the peak mass ($M_{\eta}$) and width 
($\sigma_{\eta}$) which are used to define an $\eta$ mass window for the 
given $x_F$ ($p_T$) bin.  The counts from the background function and 
signal are also used to form a relative contribution under the peak region 
from the background ($r$\,=\,$\frac{N_{\rm BG}}{N_{\rm BG}+N_{\eta}}$).

\begin{figure} [thb]
  \includegraphics[width=1.0\linewidth]{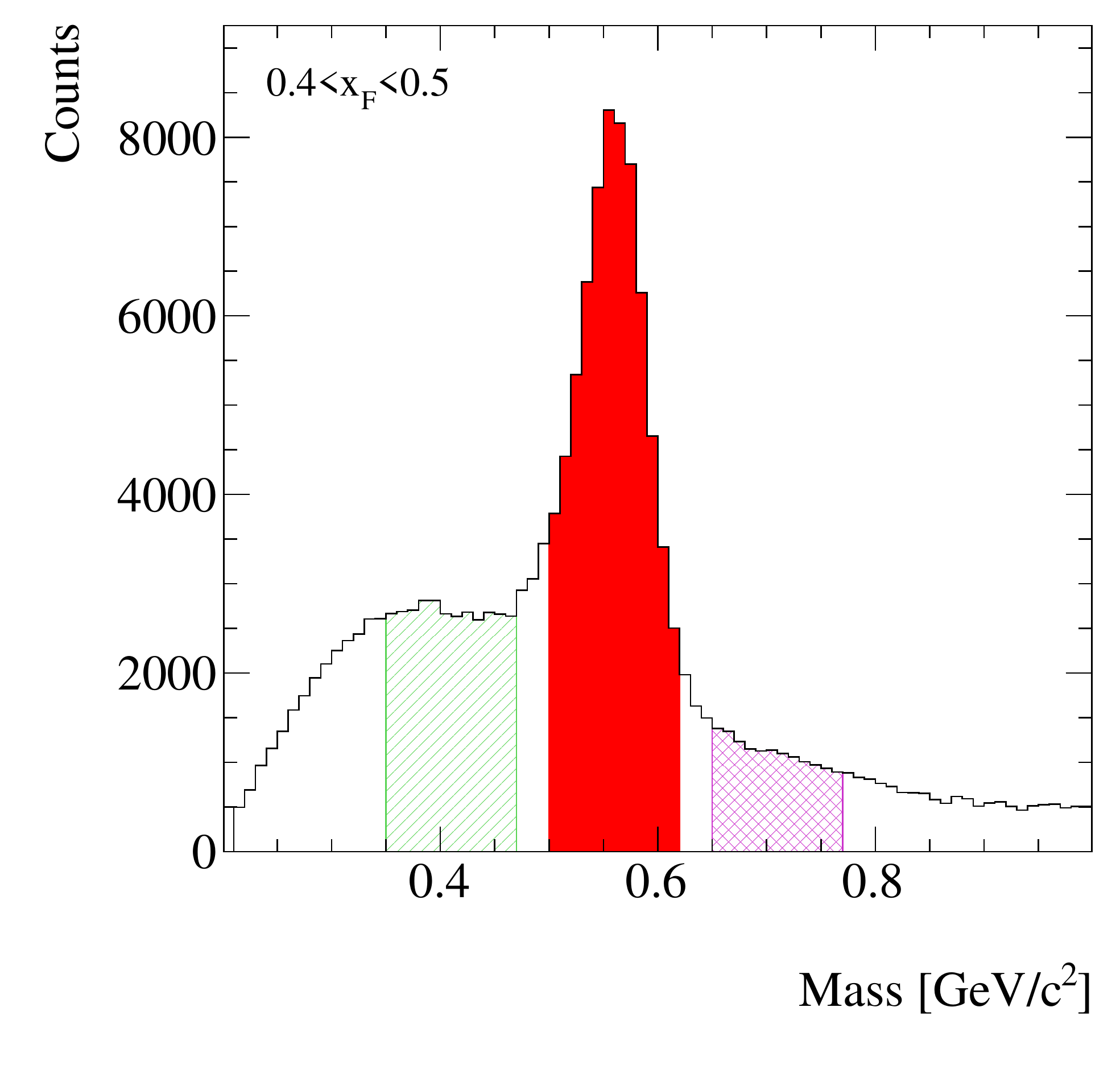}
  \caption{\label{fig:sidebands} (Color online) Invariant mass spectrum for the South
    MPC, illustrating the $\eta$ meson peak region (solid fill), as
    well as the side band regions (diagonal fill and cross-hatch).}
\end{figure}

Spin dependent and $\phi$ dependent invariant mass spectra are then 
formed, with the spin and $\phi$ dependent yields determined by 
integrating the invariant mass spectra between $M_{\eta} \pm 
2\sigma_{\eta}$.  An example of the signal and background regions are shown
in Fig.~\ref{fig:sidebands}.

\begin{figure}[thb]
  \includegraphics[width=1.0\linewidth]{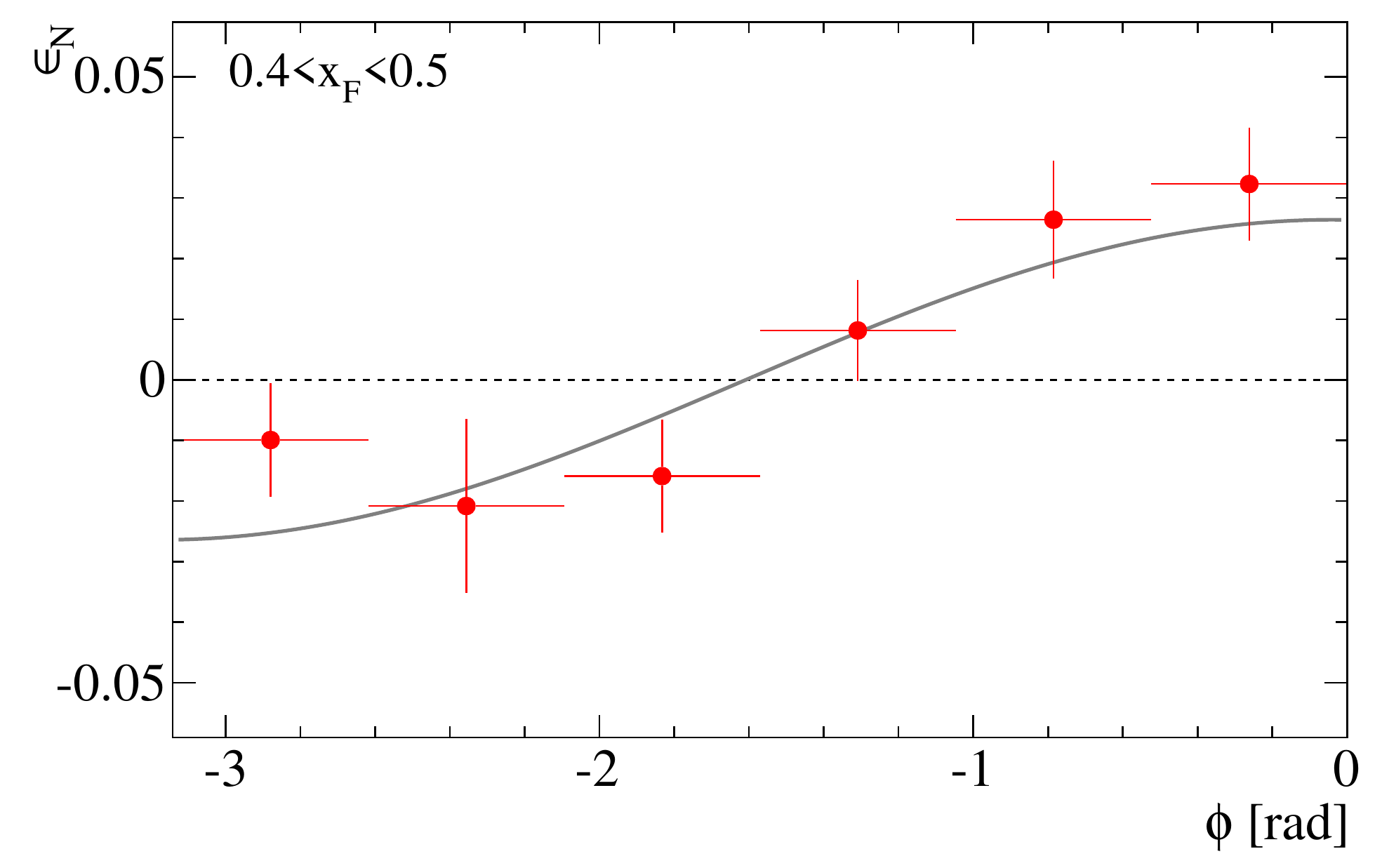}
  \caption{\label{fig:asymfit} (Color online) An example of a raw (square root method)
    asymmetry fit for a single $x_{F}$ bin in the South MPC.}
\end{figure}

The asymmetry in the peak region, $\epsilon^{M\pm 2\sigma}_N$ is then 
simply calculated from Eqs.~\ref{eq:eps_pol} and \ref{eq:eps_sqrt}. The 
resultant asymmetries are then fit with a cosine function, see 
Fig.~\ref{fig:asymfit} using the square-root formula, 
Eq.~\ref{eq:eps_sqrt}.  Note that Fig.~\ref{fig:asymfit} has six points, 
because azimuthal bins on opposite sides of the MPC are folded into each 
other when using the square-root formula.

The transverse single spin asymmetry in the $\eta$ meson peak region is 
then calculated, $A^{M\pm 2\sigma}_N$, using Eq.~\ref{eq:an_mod}.  As 
mentioned, the amplitude of the cosine function, divided by the beam 
polarization gives the value of $A^{M\pm 2\sigma}_N$. Because a 
significant background remains under the $\eta$ mass region, the final 
measurement of $A_{N}$ must be corrected for any dilution of the asymmetry 
due to this background. This background is comprised of non$\eta$ meson 
particles, which may have a different asymmetry than the signal $\eta$ 
mesons.  The correction is obtained from the asymmetry measured from a 
combined mass region from regions below 
($M_{-5\sigma}$\,$<$\,$m_{inv}$\,$<$\,$M_{-3\sigma}$~GeV/$c^2$) and above 
($M_{3\sigma}$\,$<$\,$m_{\rm inv}$\,$<$\,$M_{5\sigma}$~GeV/$c^2$) the 
$\eta$ meson mass peak, shown as the diagonal and cross-hatch filled 
regions in Fig.~\ref{fig:sidebands}, respectively. The 
background-corrected $\eta$ meson asymmetry is expressed as:

\begin{equation}
\displaystyle A^{\eta}_{N} = \frac{A^{M\pm 2\sigma}_{N} - rA^{\rm bg}_{N}}{1-r}
\label{eq:ancorr}
\end{equation}

\noindent where $r$ is the background fraction in the $\pm$2$\sigma$ 
region around the $\eta$ mass peak, $A^{M\pm 2\sigma}_{N}$ is the measured 
asymmetry of the peak region, and $A^{\rm bg}_{N}$ is the measured 
asymmetry of the background regions.  The $r$ values are found from the 
spin-independent signal and background invariant mass spectrum fits 
mentioned above.  For the lowest $x_F$ bins, calculated from the \minbias 
data, $\langle r_{\rm MB} \rangle$\,=\,0.60.  For the highest $x_F$ bins, 
calculated from the \highpt data, $\langle r_{\rm 4\times 4 B} 
\rangle$\,=\,0.37.  $A^{\rm bg}_{N}$ was found to be consistent in the low 
and high mass regions. Overall the background correction from 
Eq.~\ref{eq:ancorr} had a moderate effect of $A^{\eta}_{N}$\,$>$\,$A^{M\pm 
2\sigma}_{N}$.  A table summarizing $A^{M\pm 2\sigma}_{N}$ and $A^{\rm 
bg}_{N}$ from the \highpt triggered dataset is given in Table 
\ref{tab:taban_bkgnd_xf}.

\begin{table}[tbh]
  \caption{\label{tab:taban_bkgnd_xf} $A^{M\pm 2\sigma}_{N}$ and $A^{\rm bg}_{N}$ for $\eta$ mesons measured as a function of $x_{F}$ from the \highpt triggered dataset.  The values represented are the weighted mean of the South and North MPC. The uncertainties listed are statistical only. }
  \setlength{\tabcolsep}{3pt}
  \begin{center}
    \begin{ruledtabular} \begin{tabular}{cccccccc}
      \multirow{2}{*}{$x_{F}$ bin} & \multirow{2}{*}{$A^{M\pm2\sigma}_{N}$} & \multirow{2}{*}{Stat} & \multirow{2}{*}{$A^{\rm bg}_{N}$} & \multirow{2}{*}{Stat.}\\
          \\ \hline
          -0.7 to -0.6 & -0.0385 & 0.0602 & 0.0366 & 0.1256
          \\
          -0.6 to -0.5 & 0.0110 & 0.0186 & -0.0484 & 0.0360
          \\
          -0.5 to -0.4 & 0.0094 & 0.0094 & -0.0261 & 0.0178
          \\
          -0.4 to -0.3 & 0.0135 & 0.0117 & 0.0186 & 0.0199 
          \\ \\
          0.3 to 0.4 & 0.0314 & 0.0127 & 0.0028 & 0.0208
          \\
          0.4 to 0.5 & 0.0537 & 0.0102 & 0.0242 & 0.0190
          \\
          0.5 to 0.6 & 0.0353 & 0.0196 & 0.0458 & 0.0380 
          \\
          0.6 to 0.7 & 0.0974 & 0.0628 & 0.0147 & 0.131
          \\  
    \end{tabular} \end{ruledtabular}
  \end{center}
\end{table}

\subsection{$A_{N}$ Results}

\begin{figure}[thb]
  \includegraphics[width=1.0\linewidth]{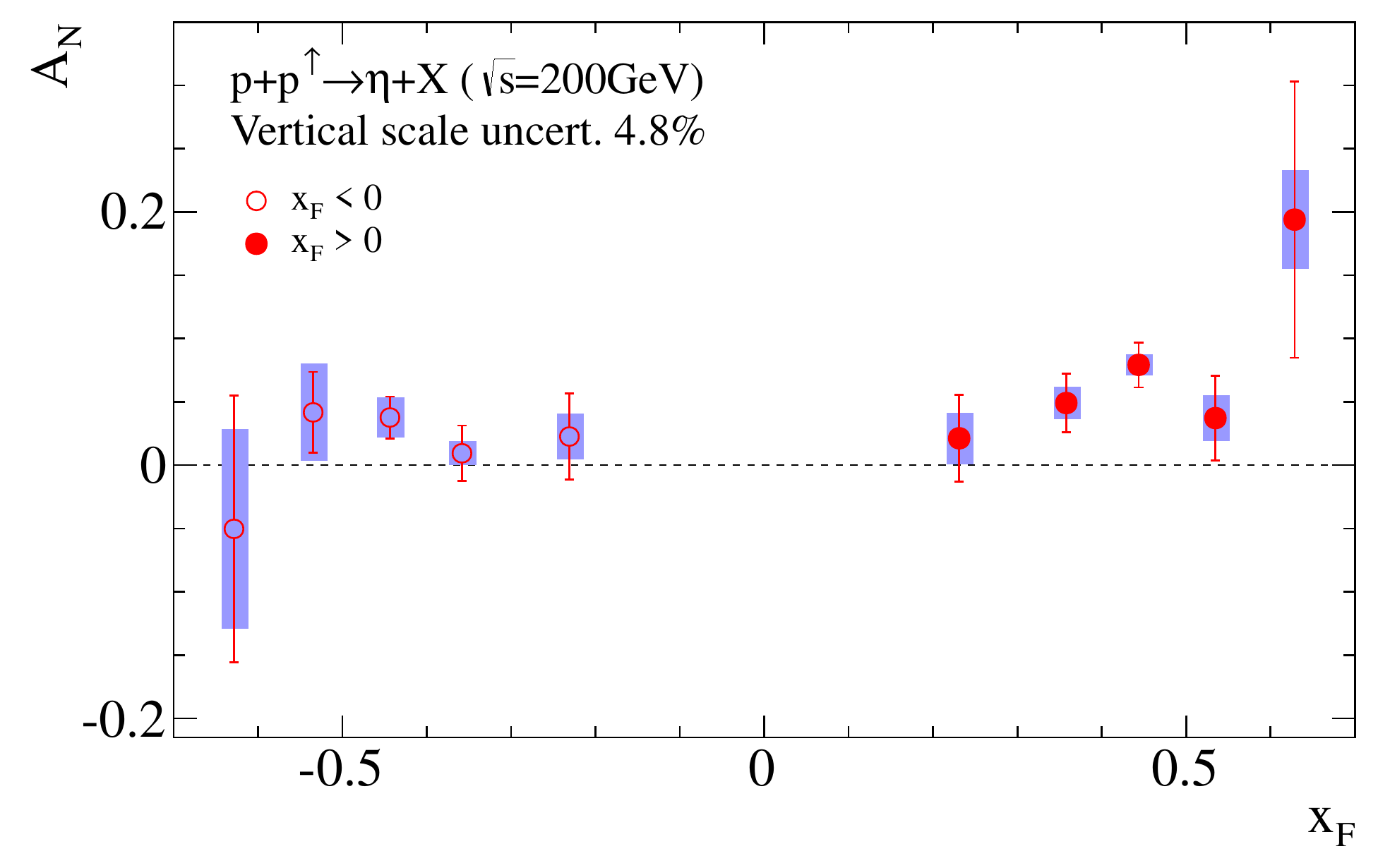}
  \caption{\label{fig:anfinal_xf}  (Color online) The $x_{F}$ dependence of $A_N$.
    The vertical error bars show the statistical uncertainty,
    the blue bands represent uncorrelated systematic uncertainties
    (see text for details).  The relative luminosity effect systematic
    uncertainties are not shown (see text and Table~\ref{tab:taban_final_xf}) }
\end{figure}

\begin{table}[thb]
  \caption{\label{tab:taban_final_xf} $A_{N}$ for $\eta$ mesons measured as
    a function of $x_{F}$.  Uncertainties listed are those due to the
    statistics, the $x_{F}$ uncorrelated uncertainties due to extracting
    the yields, and the correlated relative luminosity uncertainty
    (see text for details).}
  \setlength{\tabcolsep}{2pt}
  \begin{center}
    \begin{ruledtabular} \begin{tabular}{ccccccc}
      \multirow{2}{*}{$x_{F}$ bin} & \multirow{2}{*}{$\langle x_{F}\rangle$} & $\langle p_{T}\rangle$ & \multirow{2}{*}{$A^{\eta}_{N}$} & \multicolumn{3}{c}{Uncertainty}\\
      & & [GeV/$c$] & & stat & uncorr & corr
          \\ \hline
          -0.7 to -0.6 & -0.63 & 3.41 & -0.0503 & 0.1054 & 0.0791 & 0.0024
          \\
          -0.6 to -0.5 & -0.535 & 3.04 &  0.0417 & 0.0319 & 0.0385 & 0.0023
          \\
          -0.5 to -0.4 & -0.444 & 2.68 & 0.0376 & 0.0165 & 0.0161 & 0.0021 
          \\
          -0.4 to -0.3 & -0.358 & 2.34 & 0.0094 & 0.0219 & 0.0095 & 0.0023
          \\
          -0.3 to -0.2 & -0.231 & 1.35 & 0.0226 & 0.0339 & 0.0179 & 0.0000
          \\ \\
          0.2 to 0.3 & 0.231 &  1.35 &  0.0212 & 0.0342 & 0.0204 & 0.0000
          \\
          0.3 to 0.4 & 0.358 & 2.34 & 0.0491 & 0.0232 & 0.0127 & 0.0020
          \\
          0.4 to 0.5 & 0.444 & 2.68 & 0.0792 & 0.0177 & 0.0083 & 0.0018
          \\
          0.5 to 0.6 & 0.535 & 3.04 & 0.0372 & 0.0335 & 0.0179 & 0.0020
          \\
          0.6 to 0.7 & 0.629 & 3.41 & 0.1939 & 0.1092 & 0.0392 & 0.0019
          \\  
    \end{tabular} \end{ruledtabular}
  \end{center}
\end{table}

The $x_F$-dependent $A_N$ is shown in Fig.~\ref{fig:anfinal_xf} and 
Table~\ref{tab:taban_final_xf}, based on the weighted mean of the measured 
South and North MPC $A^{\eta}_{N}$ values. The average pseudorapidity of 
the measured $\eta$ mesons is $\langle \eta \rangle$\,=\,3.52. The 
procedure to obtain $A_N$ from the \minbias triggered dataset is the same 
as that in the \highpt dataset, and where the triggers overlap in $x_F$, 
the $A_N$ values are weighted together.  For forward $x_F$ 
($x_F$\,$>$\,0), a clear rising asymmetry is seen, ranging from 2\% to 
20\% over the measured $x_{F}$ range.  For backward $x_F$ ($x_F$\,$<$\,0), 
$A_N$ is flat and consistent with zero when averaged over $x_F$ within 
1.7$\sigma$ of the statistical plus systematic uncertainties.  An 
uncorrelated systematic uncertainty is shown as bands around the points, 
and is found by varying the functional form of the background functions.  
This changes the $M\pm 2\sigma$ range and relative $r$ values, which 
affects the number of $\eta$ mesons used in the calculation of $A_N$.  It 
also includes systematic uncertainty estimation from three different cross 
checks on the measurement of $A_N$:  increasing the mass window to 
$M$\,$\pm$\,2.5$\sigma$, the difference from the polarization formula 
measurement (Eq. \ref{eq:eps_pol}), and adding higher order cosine terms 
to the raw asymmetry fit.  The correlated systematic uncertainty (not shown on 
Fig.~\ref{fig:anfinal_xf}, see Table~\ref{tab:taban_final_xf}) is due to 
small residual relative luminosity effects in the square root formula.

\begin{figure}[thb]
  \centering
  \includegraphics[width=1.0\linewidth]{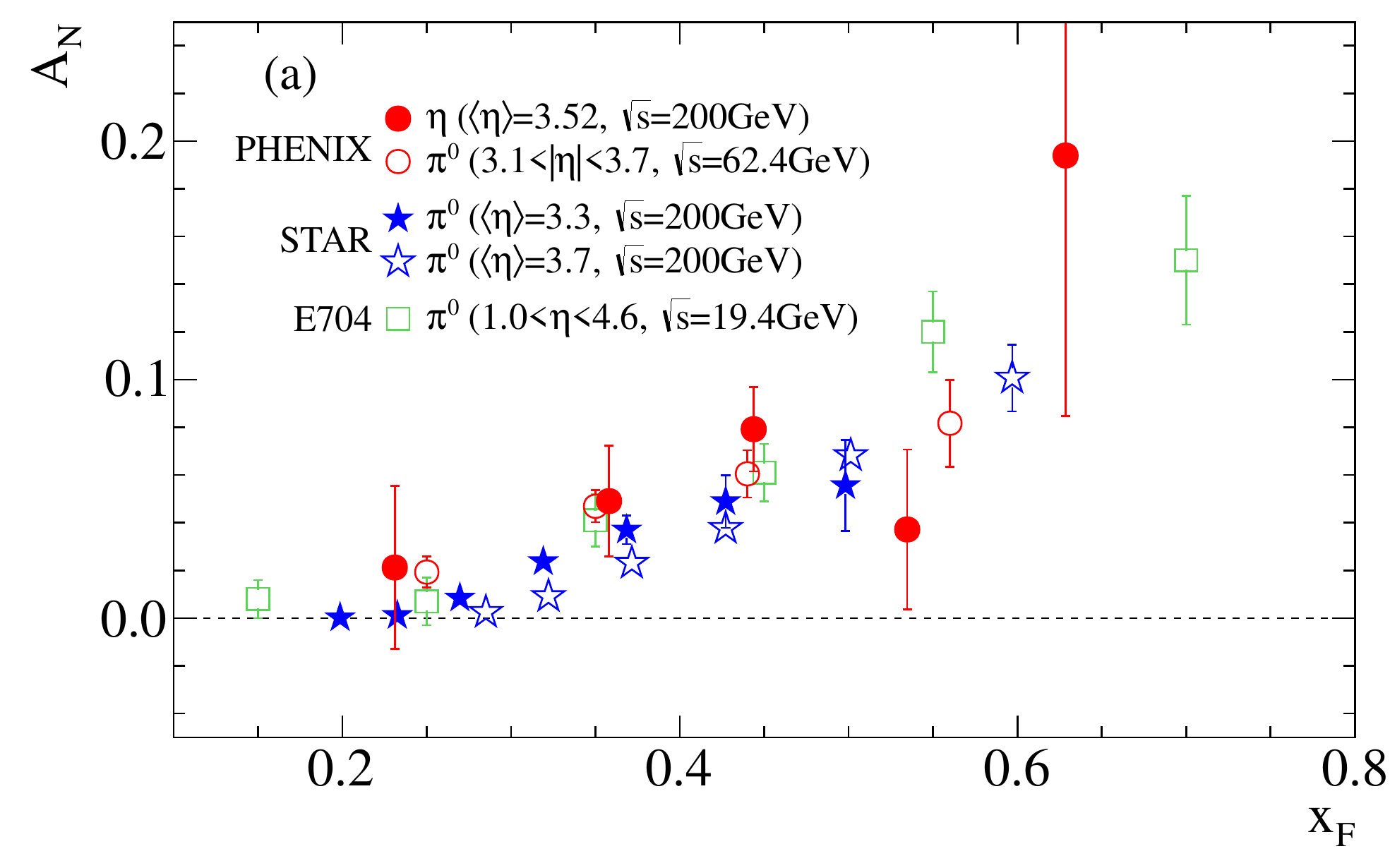}
  \includegraphics[width=1.0\linewidth]{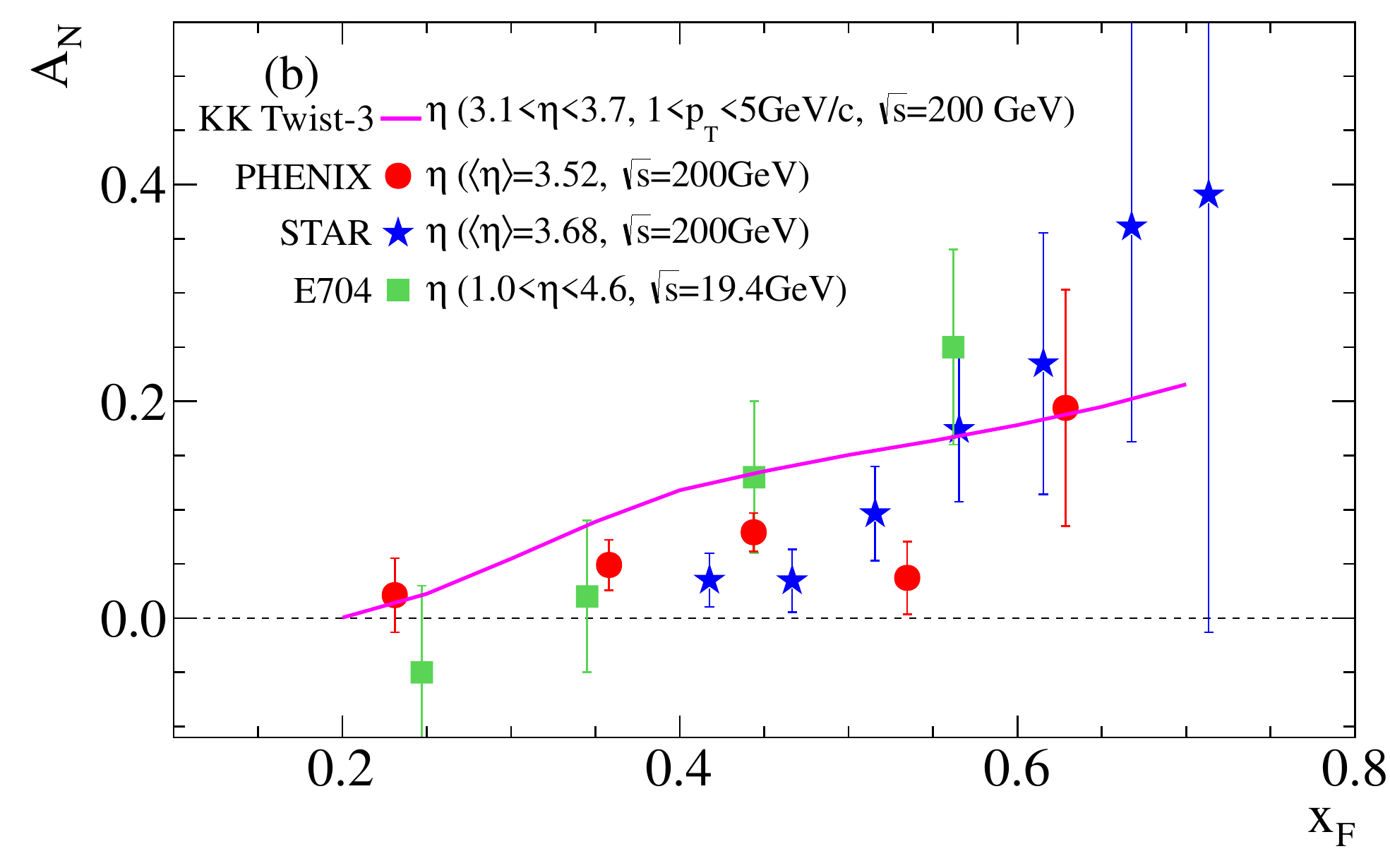}
  \caption{\label{fig:compare_etaanxf} (Color online) Comparison between the $\eta$
    meson $A_N$ and other results. Panel (a) compares with $\pi^0$ meson
    $A_N$ results from PHENIX~\cite{Adare:2013ekj}, STAR~\cite{anstar0},
    and E704~\cite{e704_0} in red/circle, blue/star, and green/square
    symbols, respectively.  Panel (b) compares to the STAR $\eta$
    meson $A_N$ result~\cite{stareta0} (blue stars), the E704 $\eta$
    meson $A_N$ result~\cite{e704eta} (green squares) and a twist-3
    calculation~\cite{kktwist33} (curve).}
\end{figure}

Figure~\ref{fig:compare_etaanxf} shows the measured $A_N$ for $\eta$ 
mesons compared to other $A_N$ measurements. The upper panel shows a 
comparison between $\eta$ meson and $\pi^0$ meson asymmetries in 
overlapping $x_F$ and similar pseudorapidity ranges at various collision 
energies. The $\eta$ meson $A_N$ is similar to the $\pi^0$ $A_N$ 
measurements at a lower center-of-mass energy made by the PHENIX 
experiment using the MPC~\cite{Adare:2013ekj}, as well as $\pi^{0}$ from 
the E704~\cite{e704_0} and STAR~\cite{anstar0} experiments.  The 
similarity between the $\eta$ and $\pi^0$ asymmetries suggests that 
initial-state spin-momentum correlations could play a role, or a common 
spin-momentum correlation is present in the fragmentation of $\pi^0$ and 
$\eta$ mesons.

The lower panel of Fig.~\ref{fig:compare_etaanxf} shows a comparison to 
measurements made by E704~\cite{e704eta} ($\sqrt{s}$\,=\,19.4\,GeV) and 
STAR~\cite{stareta0} at the same collision energy 
($\sqrt{s}$\,=\,200\,GeV). The average pseudorapidity of the PHENIX result 
is $\langle \eta \rangle$\,=\,3.52, while the average pseudorapidity of 
the STAR result is $\langle \eta \rangle$\,=\,3.68. For $x_F$\,$>$\,0.55, 
the STAR $\eta$ meson $A_N$ is larger than this PHENIX $\eta$ meson $A_N$ 
measurement, but these two results are consistent with each other within 
type-A uncertainties.

The asymmetries in Fig.~\ref{fig:compare_etaanxf} are compared to a twist-3
calculation by Kanazawa and Koike~\cite{kktwist33} based on~\cite{kktwist32},
performed for the PHENIX kinematics.  It describes the magnitude of the
asymmetry well at the lowest and highest points in $x_F$, but it is unclear
whether the observed shape for the middle $x_F$ values is well described.  No
theoretical uncertainty on the calculation is available at this time; a better
understanding of the theoretical uncertainties will be necessary in order to
draw a quantitative conclusion on the agreement with data.

\begin{figure}[thb]
  \centering
  \includegraphics[width=1.0\linewidth]{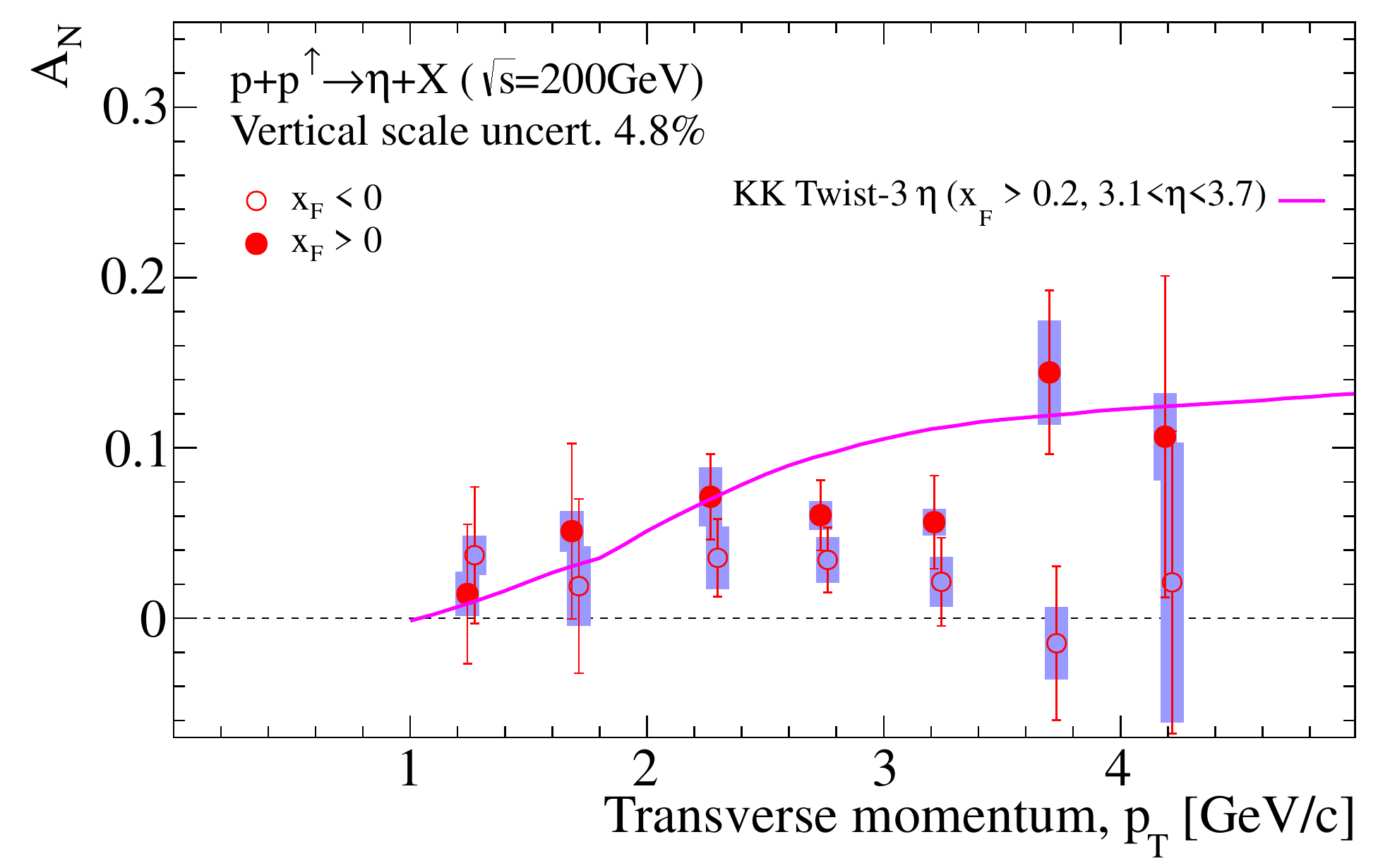}
  \caption{\label{fig:anfinal_pt}  (Color online) The $p_{T}$ dependence of $A_N$.
    The vertical error bars show the statistical uncertainty;
    the blue bands represent uncorrelated systematic uncertainties
    (see text for details).   The relative luminosity effect systematic
    uncertainties are not shown (see text and Table~\ref{tab:taban_final_pt}).
    The purple line shows a prediction
    from a twist-3 calculation based on quark-gluon correlation functions~\cite{kktwist33}.}
\end{figure}

\begin{table}[tbh]
  \caption{\label{tab:taban_final_pt} $A_{N}$ for $\eta$ mesons measured as
    a function of $p_{T}$.  Uncertainties listed are those due to the
    statistics, the $p_{T}$ uncorrelated uncertainties due to extracting
    $A_N$, and the correlated relative luminosity uncertainty
    (see text for details). }
  \setlength{\tabcolsep}{2pt}
  \begin{center}
    \begin{ruledtabular} \begin{tabular}{ccccccc}
      $p_{T}$ bin & $\langle p_{T}\rangle$ & \multirow{2}{*}{$\langle x_{F}\rangle$} & \multirow{2}{*}{$A^{\eta}_{N}$} & \multicolumn{3}{c}{Uncertainty}\\
      {[GeV/$c$]} & [GeV/$c$] & & & stat & uncorr & corr
          \\ \\
      \multicolumn{7}{c}{$x_F~<~$-0.2}
      \\ 
      1.0 to 1.5 & 1.24 & 0.23 & 0.0370 & 0.0401 & 0.0117 & 0.0000 
      \\
      1.5 to 2.0 & 1.68 & 0.27 & 0.0189 & 0.0512 & 0.0233 & 0.0000 
      \\
      2.0 to 2.5 & 2.27 & 0.42 & 0.0355 & 0.0228 & 0.0183 & 0.0042 
      \\
      2.5 to 3.0 & 2.73 & 0.44 & 0.0343 & 0.0191 & 0.0136 & 0.0041
      \\
      3.0 to 3.5 & 3.21 & 0.46 & 0.0214 & 0.0259 & 0.0149 & 0.0047
      \\
      3.5 to 4.0 & 3.70 & 0.48 & -0.0147 & 0.0452 & 0.0213 & 0.0053
      \\
      4.0 to 4.5 & 4.19 & 0.51 & 0.0211 & 0.0887 & 0.0822 & 0.0057
      \\  \\
      \multicolumn{7}{c}{$x_F~>~$0.2}
      \\ 
      1.0 to 1.5 & 1.24 & 0.23 & 0.0143 & 0.0409 & 0.0131 & 0.0000 
      \\
      1.5 to 2.0 & 1.68 & 0.27 & 0.0511 & 0.0514 & 0.0120 & 0.0000
      \\
      2.0 to 2.5 & 2.27 & 0.42 &  0.0713 & 0.0251 & 0.0176 & 0.0042 
      \\
      2.5 to 3.0 & 2.73 & 0.44 & 0.0605 & 0.0206 & 0.0085 & 0.0041 
      \\
      3.0 to 3.5 & 3.21 & 0.46 & 0.0564 & 0.0274 & 0.0078 & 0.0047
      \\
      3.5 to 4.0 & 3.70 & 0.48 & 0.1443 & 0.0480 & 0.0306 & 0.0053
      \\
      4.0 to 4.5 & 4.19 & 0.51 & 0.1066 & 0.0944 & 0.0257 & 0.0057
      \\ 
    \end{tabular} \end{ruledtabular}
  \end{center}
\end{table}

The $p_T$ dependence of the asymmetry is shown in 
Fig.~\ref{fig:anfinal_pt} and Table~\ref{tab:taban_final_pt}. For $A_N$ 
measured at forward $x_F$ ($x_F$\,$>$\,0.2), a clear nonzero asymmetry is 
seen ($\langle A_{N} \rangle$\,=\,0.061\,$\pm$\,0.012), while $A_N$ for 
backward $x_F$ ($x_F$\,$<$\,-0.2) is consistent with zero within 
1.7$\sigma$.  The uncorrelated and correlated systematic uncertainties are 
evaluated the same way as in the $x_F$ dependence of $A_N$.

Figure~\ref{fig:anfinal_pt} also shows the measured $A_N$ as a function of 
$p_T$ compared to the twist-3 calculations. Similar to the case for the 
$x_F$ dependence, the twist-3 calculation describes the magnitude of the 
asymmetry well at the lowest and highest measured points in $p_T$, but it 
is not clear if it describes the observed shape in the mid-$p_T$ range. It 
should be noted that the data points in $p_T$ are integrated over a wide 
range of $x_F$, 0.2\,$<$\,$x_F$\,$<$\,0.7.

\section{Summary}

By utilizing data taken by the MPC detector installed at forward rapidity 
in the PHENIX experiment at RHIC, the invariant cross section as a 
function of $p_T$ and the transverse single spin asymmetry $A_N$ as a 
function of $x_F$ and $p_T$ have been measured for inclusive $\eta$ mesons 
produced at forward rapidity ($\langle \eta \rangle$\,=\,3.52) from 
$p^{\uparrow}+p$ collisions at a center-of-mass energy of 
$\sqrt{s}$\,=\,200~GeV.  The NLO pQCD calculation was found to be 
consistent with the invariant cross section measurement at momentum of 
$p_T$\,$>$\,1.5~GeV/$c$. This measurement can be used to improve 
constraints on the hadronization process of $\eta$ mesons in future global 
analyses of the $\eta$ fragmentation function.  Non-zero asymmetries 
measured at forward $x_F$ are consistent with previous $\pi^{0}$ meson 
results within statistical uncertainties. Because the $\pi^0$ and $\eta$ 
mesons are produced from potentially different parton fractions, and also 
might have different polarized fragmentation functions due to isospin or 
mass differences or the presence of strange quarks in the $\eta$, this 
data will help to constrain the relative importance of spin-momentum 
correlations in the initial state polarized protons versus that of 
spin-momentum correlations in the fragmentation. The dependencies of the 
measured asymmetry on $x_F$ and $p_T$ are reasonably well described by 
twist-3 calculations using quark-gluon correlation functions; a 
quantitative comparison can be made once uncertainties become available on 
the calculations. With higher statistics from future data sets, a doubly 
differential measurement of the asymmetry binned in both $x_F$ and $p_T$ 
simultaneously could provide a much more stringent test of any available 
calculations and better constrain twist-3 quark-gluon correlation 
functions if they turn out to be the dominant contribution.


\section*{ACKNOWLEDGMENTS}     

We thank the staff of the Collider-Accelerator and Physics
Departments at Brookhaven National Laboratory and the staff of
the other PHENIX participating institutions for their vital
contributions.  We acknowledge support from the 
Office of Nuclear Physics in the
Office of Science of the Department of Energy,
the National Science Foundation, 
Abilene Christian University Research Council, 
Research Foundation of SUNY, and
Dean of the College of Arts and Sciences, Vanderbilt University 
(U.S.A),
Ministry of Education, Culture, Sports, Science, and Technology
and the Japan Society for the Promotion of Science (Japan),
Conselho Nacional de Desenvolvimento Cient\'{\i}fico e
Tecnol{\'o}gico and Funda\c c{\~a}o de Amparo {\`a} Pesquisa do
Estado de S{\~a}o Paulo (Brazil),
Natural Science Foundation of China (P.~R.~China),
Ministry of Science, Education, and Sports (Croatia),
Ministry of Education, Youth and Sports (Czech Republic),
Centre National de la Recherche Scientifique, Commissariat
{\`a} l'{\'E}nergie Atomique, and Institut National de Physique
Nucl{\'e}aire et de Physique des Particules (France),
Bundesministerium f\"ur Bildung und Forschung, Deutscher
Akademischer Austausch Dienst, and Alexander von Humboldt Stiftung (Germany),
OTKA NK 101 428 grant and the Ch. Simonyi Fund (Hungary),
Department of Atomic Energy and Department of Science and Technology (India), 
Israel Science Foundation (Israel), 
National Research Foundation of Korea of the Ministry of Science,
ICT, and Future Planning (Korea),
Physics Department, Lahore University of Management Sciences (Pakistan),
Ministry of Education and Science, Russian Academy of Sciences,
Federal Agency of Atomic Energy (Russia),
VR and Wallenberg Foundation (Sweden), 
the U.S. Civilian Research and Development Foundation for the
Independent States of the Former Soviet Union, 
the Hungarian American Enterprise Scholarship Fund,
and the US-Israel Binational Science Foundation.



\end{document}